\input harvmac.tex

\input epsf

\lref\mikh{Mikhailov A.V.: The reduction problem and the inverse 
scattering method. Physica {\bf D3}, 73 (1981) }

\lref\olsh{Mikhailov A.V., Olshanetskii M.A., Perelomov A.M.:
Two-dimensional generalized Toda lattice. Commun.Math.Phys. {\bf 
179}, 401 (1981)}

\lref\verg{Vergeles S., Gryanik V.:Two dimensional quantum field 
theories having exact solutions. Yad.Fiz. {\bf 23}, 1324  (1976)}

\lref\zamzam{Zamolodchikov A.B.: Quantum Sine-Gordon model. The total S 
matrix. ITEP-12-1977, 1977}

\lref\ari{Arinshtein A.E., Fateev V.A., Zamolodchikov A.B.: Quantum 
S matrix of the (1+1)-dimensional Toda chain. Phys.Lett. {\bf B87}, 
389 (1979)}

\lref\fatzam{Fateev V.A., Zamolodchikov A.B.: Conformal field theory 
and purely elastic S matrices. Int.J.Mod.Phys.{\bf A5}, 1025  (1990) }

\lref\cor{Braden H.W., Corrigan E., Dorey P.E., Sasaki R.:
 Affine Toda field theory and exact S matrices. Nucl.Phys. {\bf B338}, 
689 (1990) }

\lref\mus{Christe P., Mussardo G.: Elastic S matrices 
in(1+1)-dimensions and Toda field theories. Int.J.Mod.Phys.{\bf A5}, 
4581  (1990)}

\lref\muss{Christe P., Mussardo G.: Integrable systems away from 
criticality: the Toda field theory and  S matrix of the tricritical 
Ising model. Nucl.Phys. {\bf B330}, 465 (1990)}

\lref\ser{Lukyanov S.: Free field representation for massive 
integrable models.   Commun.   Math.  Phys.    {\bf 167},183  (1995) }

\lref\fei{Feigin B., Frenkel E.: Quantum W algebras and elliptic 
algebras. Commun.Math.Phys.  {\bf 178}, 653  (1996)}

\lref\sero{Lukyanov S.: Correlators of the Jost functions in the 
Sine-Gordon model. Phys.Lett. {\bf B325}, 409  (1994)} 

\lref\zam{Zamolodchikov A.B.: Unpublished}

\lref\serr{Lukyanov S.: Form factors of exponential fields in the 
Sine Gordon  model.  Mod. Phys. Lett.    {\bf A12},  2543   (1997) }

\lref\sert{Lukyanov S.: Form factors of exponential fields in the 
affine $A_{N-1}^{(1)}$ Toda model. Phys.Lett. {\bf B408}, 192 (1997) }

\lref\kar{Berg B.,  Karowski M.,  Weisz P.: Construction of Green 
functions from an exact S matrix. Phys.Rev. {\bf D19}, 2477  (1979) }

\lref\ale{Zamolodchikov Al.B.: Two point correlation function in 
scaling Lee-Yang model. Nucl.Phys. {\bf B348}, 619  (1991) }

\lref\frmussim{Fring A., Mussardo G.,  Simonetti P.: Form factors for 
integrable lagrangian field theories, the Sinh-Gordon 
model. Nucl.Phys. {\bf B393}, 413  (1993)}

\lref\frmussimm{Fring A., Mussardo G.,  Simonetti P.: Form factors of 
the elementary field in the 
Bullough-Dodd model. Phys.Lett. {\bf B307}, 83  (1993)}

\lref\delsica{Delfino G., Simonetti P., Cardy J.L.: Asymptotic 
factorization of form factors in two dimensional quantum field theory. 
Phys. Lett. {\bf B387}, 327  (1996) }

\lref\oo{Oota T.: Functional equations of form factors for diagonal 
scattering theories. Nucl.Phys. {\bf B466}, 361 (1996) }

\lref\ooo{Oota T.: Q deformed Coxeter element in nonsimply laced 
affine Toda field theories. Nucl.Phys. {\bf B504}, 738 (1997) }

\lref\pil{Pillin M.: Polynomial recursion equations  in form factors 
of $A-D-E$ Toda field theories. Lett.Math.Phys.{\bf 43}, 211  (1998)}

\lref\wa{Watson K.M.: Phys.Rev. {\bf 95}, 228 (1954)}

\lref\karow{Karowski M, Weisz P.: Exact form factors in $ 
(1+1)$-dimensional field theoretic models with soliton behavior. 
Nucl.Phys. {\bf B139}, 455  (1978) }

\lref\fed{Smirnov F.A.: Form factors in completely integrable models 
of quantum field theory. Singapore: World Scientific (1992) 208 p. (Advanced series
in mathematical physics, 14). }

\lref\fe{Smirnov F.: The quantum Gelfand-Levitan-Marchenko equations
and form factors in the Sine-Gordon model.J.Phys. {\bf A17}, L873  (1984)}

\lref\ddv{Destri C., de Vega H.J.: New exact results in affine Toda 
field theories: free energy and wave function renormalization. 
Nucl.Phys. {\bf B358}, 251  (1991) }

\lref\br{Brazhnikov V., Lukyanov S.: Angular quantization and form 
factors in massive integrable models.Nucl.Phys. {\bf B512}, 616 (1998) }

\lref\resh{Frenkel E., Reshetikhin N.: Quantum Affine Algebras and 
Deformations  of the
 Virasoro and W-algebras. {\bf q-alg/9505025}}
 
\lref\reshh{Frenkel E., Reshetikhin N.: Deformations of W-algebras associated 
to simple Lie algebras. {\bf q-alg/9708006}}

\lref\jim{Jimbo J., Miwa T.: Algebraic analysis of Solvable Lattice 
Models. Kyoto Univ., RIMS-981 (1994)}

\lref\foda{Davies B., Foda O., Jimbo J., Miwa T., Nakayashiki A.: 
Diagonalization of the $XXZ$ Hamiltonian by vertex operators. Commun. 
Math. Phys. {\bf 151}, 89 (1993)}

\lref\servir{Lukyanov S.: A note on the deformed Virasoro 
algebra. Phys.Lett. {\bf B367},121 (1996) }

\lref\bax{Baxter R.J.: Exactly solved models in statistical 
mechanics. London: Academic Press 1982}

\lref\acer{Acerbi C.: Form factors of exponential operators and  exact 
wave function renormalization constant in the  Bullough-Dodd model. 
 Nucl.Phys. {\bf B497}, 589  (1997)}
 
\lref\slava{Lukyanov S., Pugai Ya.: Bosonization of ZF algebras: direction toward
 deformed Virasoro algebra. J.Exp.Theor.Phys. {\bf 82}, 1021 (1996)}
 
\lref\pilch{Bouwknegt P.,  Pilch K.: On deformed W-algebras and 
quantum affine algebras. {\bf q-alg/9801112}}

\lref\ku{ Kuniba A., Suzuki J.: Analytic Bethe ansatz for fundamental 
representations of Yangians. Commun.Math.Phys.{\bf 173}, 225 (1995) }
 
\lref\fat{Fateev V.A.: To be published}

\lref\clava{Clavelli L., Shapiro J.A.: Pomeron factorization in 
general dual model. Nucl. Phys. {\bf B57}, 490  (1973)}

\lref\fate{Fateev V.A.: Unpublished.} 

\Title{\vbox{\baselineskip12pt\hbox{UMTG-212}
                              \hbox{hep-th/9809074}}}
{\vbox{\centerline{
Wave function renormalization constants and }
\vskip6pt
\centerline{ one-particle form  factors in}
\vskip6pt
\centerline{ $D_{l}^{(1)}$ Toda field theories.}}}
\vskip15pt
\centerline{Vadim A. Brazhnikov \footnote{$^*$}{e-mail address:
vadim@physics.miami.edu }}
\centerline{}
\centerline{Department of Physics, The University of Miami }
\centerline{James L. Knight Physics building} 
\centerline{1320 Campo Sano Drive}
\centerline{Coral Gables, Fl 33146, USA}
\bigskip\bigskip\bigskip
\centerline{{\bf Abstract}}
We apply the method of angular quantization to calculation of the 
wave function renormalization constants in $D_{l}^{(1)}$ affine Toda  
quantum field theories. A general formula for the wave function 
renormalization  constants in ADE Toda field theories is proposed.  We 
also  calculate all one-particle form factors 
and some of the two-particle form factors  of an exponential field.
\bigskip
\Date{August, 1998}
\vfill
\eject

\newsec{Introduction}

Two dimensional affine Toda quantum field theories (ATQFT) have attracted 
a great deal of attention in recent years. These theories can be 
associated to any simple Lie algebra $\cal G$ of rank $l$ \mikh 
\olsh \ . Let 
$\{{\vec \alpha}_{k}\}_{k=1}^{l}$ be a set of positive simple roots of the 
Lie algebra $\cal G$,  ${\bf C}_{p q}$ is its Cartan matrix
\eqn\cart{{\bf C}_{p q}\,=\,{{2 \,{\vec \alpha}_{p}\cdot{\vec 
\alpha}_{q}}\over{{\vec \alpha}_{p}\cdot{\vec \alpha}_{p}}}\,\, .}
and  ${\vec \alpha}_{0}$ is the negative of the highest root
\eqn\maxroot{-{\vec \alpha}_{0}\,=\,\sum_{k=1}^{l}\,n_{k}\,{\vec 
\alpha}_{k}\,\, .}
The affine $\cal G$ Toda model describes the dynamics of an
$l$-component real scalar field
\eqn\field{{\vec \varphi}(x)\,=\,\left( \varphi^{(1)}(x), \ldots , 
\varphi^{(l)}(x) \right)\,\, .}
governed by the Euclidean action
\eqn\action{ {\cal A}\,=\,\int d^2 x \,
\bigg\{\, {1\over {8 \pi}} \big(\partial_{\nu}{\vec \varphi} \big)^2 +
\mu\, \sum_{k=0}^{l}\, n_{k}\, e^{b {\vec \alpha}_{k} {\vec \varphi} } 
\, \bigg\}\,\, .}
where the integers $n_{k},\, k=1,\ldots ,l$ are  defined by \ 
\maxroot \ , $n_{0} = 1$ and $b$ is a real coupling constant.
The affine Toda field theories  are completely integrable at the 
classical level \mikh \olsh . The quantum integrability of the ATQFT 
is best understood for the simply laced algebras, though the exact  
$S$-matrices were proposed for all simple Lie algebras  \verg \zamzam 
\ari \fatzam \cor \mus \muss .

An important feature of the  ATQFTs based on simply laced algebras is that 
the  classical mass ratios are unaffected by renormalization . Another important 
property of the 
of these models is that the $S$-matrices are invariant under the duality 
transformation   $b \rightarrow b^{-1}$. A possible way to study off-shell 
properties of ATQFTs \kar \ale \frmussimm \frmussim \oo \pil \
is provided by a form factor approach originated in the works \wa \verg 
\karow \fe .

Recently a new powerful method has been designed \ser ,\sero ,\slava \ for studying integrable 
two dimensional models. The key point of this method is to use angular 
quantization of the theories \zam . In this approach the representation of the algebra 
of local fields is associated with a half-infinite line. The space of 
representation is usually called the angular quantization space. 
The roots of this approach go back to Baxter's corner matrix method 
\bax \jim \foda . The angular quantization space  is a 
field-theoretical analog of the space where the lattice corner 
transfer matrix acts, and inherits the remarkable features of the 
latter. Using this method, the form factors of exponential operators 
and  the wave functions renormalization constants for  $A_{l}^{(1)}$ 
ATQFT were found in \serr \sert . The renormalization constants coincide 
with the ones calculated in \ddv . The method was also successfully 
applied \br \ to the calculation of form factors in $A_{2}^{(2)}$ ATQFT 
\acer .  It was 
proposed in \servir  \br \  that the angular quantization space 
for a massive integrable model can be treated as a ``scaling`` limit 
of a representation of some ``deformed`` algebra. In this limit 
the currents of the deformed algebra become Zamolodchikov-Faddeev (ZF)
operators corresponding to a diagonal scattering theory. In  
\resh \br \reshh \ a striking similarity of the free field 
representations of ZF operators and  the Baxter $T\,-\,Q$ equation 
was observed.

In this paper we will deal with $D_{l}^{(1)}$ ATQFT. Using a free field 
representation of the angular quantization space  we obtain 
all one-particle and some of the two-particle  form factors of an 
exponential  operator $e^{{\vec a}{\vec \varphi}}$. This 
provides enough information to calculate the wave function renormalization 
constants $Z_{k}, k = 1,\ldots, l$, which define residues of the two-point 
functions of fields $\phi^{(k)}$ that diagonalize the mass matrix of \ \action \
\eqn\res{\langle\,vac\,|\, \phi^{(k)}(p)\, \phi^{(k)}(-p) \,|\,vac\,\rangle 
\,\rightarrow\,{{4 \pi i  Z_{k}}\over{p^{2} - m^{2}_{k} + i \epsilon}}\,\, ,\hskip40pt 
p^{2}  \rightarrow m_{k}^{2}\,\, .}
where $m_{k}$ is the mass of the $k$-th particle $(2.8)$ of the theory. 
The expression  for $Z_{k}$ is given by
\eqn\mainf{\eqalign{&Z_{k}\,=\,\exp \bigg\{ -4 \int\limits_{0}^{\infty} {{d \nu} \over {\nu}}    
\, \sinh({{\pi b \nu}\over{h Q}}) \sinh({{\pi 
\nu}\over{h Q b}}) \,\left(\,\coth(\pi \nu) \,\bigl(C^{-1}(\nu)\bigr)_{k 
k}\,-\, e^{-{{\pi \nu}\over{h}}}\,\right) \bigg\} \,\, ,\cr
&\hskip265pt k = 1,\ldots, l\,\, .}} 
where $Q = b + b^{-1}$ and $h = 2 (l-1)$ is the dual Coxeter number. 
The matrix $C$ is a ``deformed`` Cartan matrix
\eqn\defcar{C_{p q}(\nu)\,=\,4 \sinh^{2}({{\pi \nu}\over{2 h}}) 
\,\delta_{p q}\,+\,{\bf C}_{ p q }\,\, .}
The renormalization constants for $A_{l}^{(1)}$ Toda model \ddv \ can also be 
written in the form  \ \mainf \ where the  matrix $C$ is the ``deformed`` 
$A_{l}$  Cartan matrix . We expect that the formula 
\ \mainf \ is also valid for $E_{l}^{(1)}$ ATQFT. 

The exact result  for $Z_{k}$ \ \mainf \  satisfy necessary 
field-theoretical requirements. In particular, $0 < Z_{k} \le 1 $, 
and it matches the one loop perturbative check.

\vskip 6pt
\centerline{\epsfxsize=3truein\epsfbox{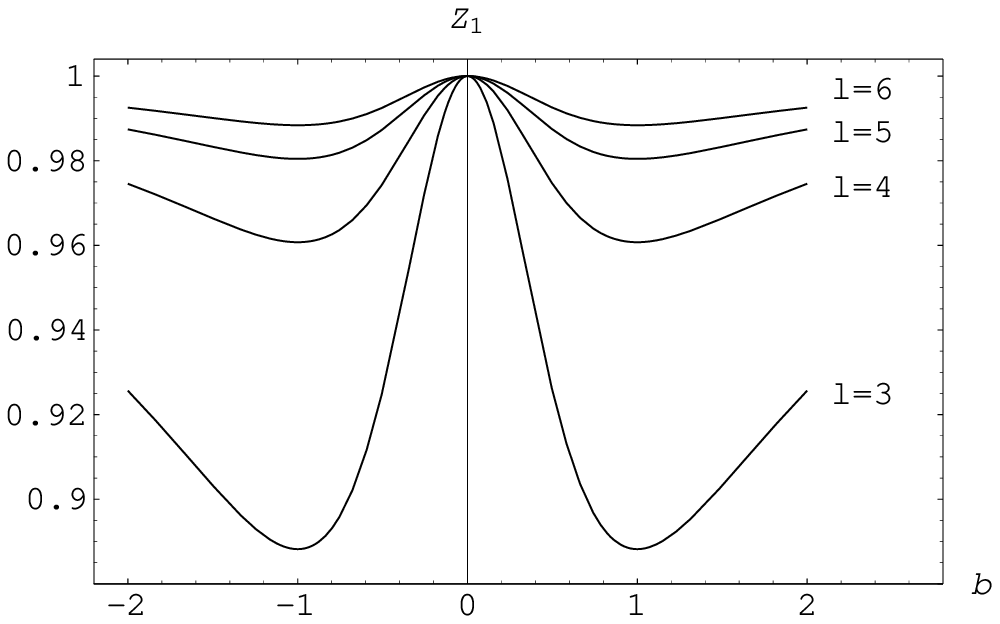}\hfil%
\epsfxsize=3truein\epsfbox{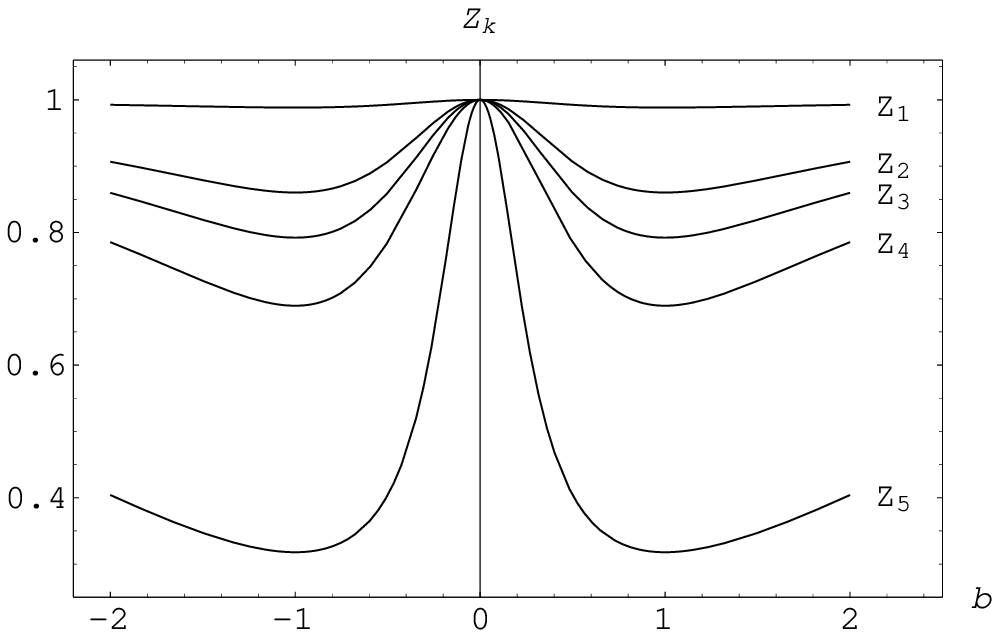}}
\vskip 3pt
\hskip1.1truein Fig.$\,1$ \hskip2.8truein Fig.$\,2$
\vskip 6pt

A plot of the functions $Z_{1}(b)$ for the first few $D_{l}^{(1)}$ 
Toda theories is given in Figure 1.  The Figure 2 shows the functions  
$Z_{k}(b)$ for $D_{6}^{(1)}$ Toda theory. Note, that while $Z_{1}(b)$ 
is very close to $1$, the deviation of $Z_{5}(b) = Z_{6}(b)$ from $1$
is considerable.

Here is the layout of the paper.  In Section 2 we introduce basic 
notations and facts about the $D_{l}^{(1)}$ Toda model. In Section 3 the 
method of reconstruction of form factors (the angular quantization 
method) is briefly described. In Section 4 we construct the free field 
representation for form factors and test it in Section 5 on the example of 
$D_{4}^{(1)}$ ATQFT. Section 6 contains the main results of the 
paper - the one-particle form factors and the wave function renormalization 
constants. Finally, we conclude with general remarks in Section 7.

\newsec{$S$-matrix and spectrum of $D_{l}^{(1)}$ Toda field theory}

The affine $D_{l}^{(1)}$ Toda model describes the dynamics of an
$l$-component real scalar field ${\vec \varphi}(x)$ \ \field \
governed by the Euclidean action \ \action \ .
The vectors ${\vec \alpha}_{k}$, $k\,=\,1,\ldots,l$ 
are the simple positive roots of the Lie algebra $D_{l}$, ${\vec 
\alpha}_{k}^{2} = 2$.
The integers $n_{k}\,=\,1$ if $k\,=\,0,\,1,\,l,\,l-1$ and 
$n_{k}\,=\,2\,$ otherwise. In terms of an orthonormal basis in ${\bf R}^{l}$
\eqn\orbasis{ {\vec \varepsilon}_{a}\cdot{\vec 
\varepsilon}_{b}\,=\,\delta_{{a b}}\, ,\hskip30pt 
a,b\,=\,1,\ldots,l\,\, ,}
the simple positive roots can be expressed as 
\eqn\rootsbas{\eqalign{&{\vec \alpha}_{k}\,=\,{\vec \varepsilon}_{k}-{\vec 
\varepsilon}_{k+1}\, ,\hskip30pt k\,=\,1, \ldots l-1\, ,\cr
&{\vec \alpha}_{l}\,=\,{\vec \varepsilon}_{l-1} + {\vec 
\varepsilon}_{l}\, .}}
The lagrangian \ \action \  possesses 
explicit symmetry under the action of a group $G\,=\,Z_{2} \times Z_{2}$ 
for even $l$ or $G\,=\,Z_{4}$ if $l$ is odd. It is convenient to 
introduce  new fields \cor \ $\phi^{(k)}(x)$, $k\,=\,1 ,\ldots ,l$
corresponding to the irreducible representations of the group $G$:
\eqn\newfield{\varphi^{(k)}\,=\,2 
h^{-{{1}\over{2}}}\,\sum_{p=1}^{l-2}\, \sin({{2 \pi p 
(k-1)}\over{h}})\,\phi^{(p)},\,\,\,\,\,\,\,\,k\, \neq 1,l\,\, ,}
for even $l$ we put 
\eqn\newfieldev{\varphi^{(1)}\,=\,{{\phi^{(l)}\,+\,\phi^{(l-1)}}\over 
{\sqrt 2}},
\,\,\,\,\,\,\,\,\varphi^{(l)}\,=\,(-1)^{{{l}\over{2}}} \,\, 
{{\phi^{(l)}\,-\,\phi^{(l-1)}}\over {\sqrt 2}}\,\, ,}
for odd $l$ we put 
\eqn\newfieldod{\varphi^{(1)}\,=\,{{\phi^{(l)}\,+\,\phi^{(l-1)}}\over 
{{\sqrt 2}}},
\,\,\,\,\,\,\,\,\varphi^{(l)}\,=\,(-1)^{{{l+1}\over{2}}}\,\, 
{{\phi^{(l)}\,-\,\phi^{(l-1)}}\over {i\,{\sqrt 2}}}\,\, .}
In terms of these new fields the mass matrix for the lagrangian \ 
\action \ becomes diagonal. The action of the group $G$  for even $l$ is 
generated by the  group elements $g_{1}$ and $g_{2}$, \ \ $g_{1}^{2}=g_{2}^{2}\,=\,1$
\eqn\groupactionev{\eqalign{& g_{1}:\,\,\,\,\,\,\phi^{(k)}\,\rightarrow\,(-1)^{k} 
\phi^{(k)},\hskip85pt k\,=\,1, \ldots ,l-2\,\, ,\cr
&\hskip27pt\phi^{(l-1)}\,\rightarrow\, 
- (-1)^{{{l}\over{2}}}\,\phi^{(l-1)},\,\,\,\,\,\,\phi^{(l)}\,\rightarrow\,
(-1)^{{{l}\over{2}}}\,\phi^{(l)}\,\, .\cr
& g_{2}:\,\,\,\,\,\,\phi^{(k)}\,\rightarrow\,\phi^{(k)},\hskip110pt k\,=
\,1, \ldots ,l-2\,\, ,\cr
&\hskip27pt \phi^{(l-1)}\,\rightarrow 
\,-\,\phi^{(l-1)},\,\,\,\,\,\,\phi^{(l)}\,\rightarrow \,-\,\phi^{(l)}\,\, .\cr}}
For odd $l$ the action of the group $G$ is generated by the element 
$g$, \ \ $g^{4}\,=\,1$
\eqn\groupactionod{\eqalign{ g:\,\,\,\,\,\,&\phi^{(k)}\,\rightarrow\,(-1)^{k} 
\phi^{(k)},\hskip80pt k\,=\,1, \ldots ,l-2\,\, ,\cr
&\phi^{(l-1)}\,\rightarrow\,-\,i^{l}\, 
\phi^{(l-1)},\,\,\,\,\,\,\phi^{(l)}\,\rightarrow\, i^{l} \,
\phi^{(l)}\cr}}

The spectrum of the $D_{l}^{(1)}$ ATQFT consists of $l$ 
particles $\{ B_{k}\}_{k=1}^{l}$. These particles  are in one-to-one 
correspondence with fundamental representations $\{ 
\pi_{k}\}_{k=1}^{l}$  of the Lie algebra 
$D_{l}$ and their masses are given by
\eqn\masses{\eqalign{&m_{k}\,=\,2 m\,\sin ({{k \pi}\over{h}}),\,\,\,\,\,\,\,
k\,=\,1,\ldots,l-2 \,\, ,\cr
&m_{l-1}\,=\,m_{l}\,=\,m \,\, .\cr}}
where $h\,=\,2 (l-1)$ is the dual Coxeter number. A linear basis in the 
physical Hilbert space  $\pi_{A}$ of the theory is provided by a set 
of asymptotic states\foot{Our convention for the normalization of the asymptotic states is
$$ \langle\, vac\, |\, vac\, \rangle=1\,, \ \ \ \ \ \ \
\langle\,   B_{p}(\theta)\, |\, B_{k}(\theta')
\, \rangle=2\pi\ \delta_{p k} \delta(\theta-\theta')\ .$$}
\eqn\asymp{|\, B_{k_{1}}(\theta_1)...B_{k_{n}}(\theta_n)\, \rangle\ ,}
where rapidities are ordered as $\theta_1>...> \theta_n$.

The two-particle $S$-matrix, describing $B_{a} B_{b} \rightarrow B_{a} 
B_{b}$ scattering was proposed in \cor . In a compact form it can be 
written \ooo , \reshh \ as 
\eqn\intsmat{S_{a b}(\theta)\,=\,\exp \bigg\{ -4 \int\limits_{-\infty}^{\infty} {{d \nu} \over {\nu}}    
\,e^{i \nu \theta} \sinh({{\pi b \nu}\over{h Q}}) \sinh({{\pi 
\nu}\over{h Q b}}) \,(C^{-1}(\nu))_{a b}\,\bigg\} \,\, , }
where the matrix $C$ was defined in \ \defcar \
To analyze the  analytical 
structure of the two-particle $S$-matrix it is useful to present its 
matrix elements as a product of meromorphic functions, 
\eqn\ampl{\eqalign{&S_{a b}(\theta)\,=\,\prod_{p=1}^{min(a,b)}\,{{{\cal 
F} (\theta -{{i \pi}\over{h}} (a+b+1-2 p))}\over {{\cal 
F} (\theta +{{i \pi}\over{h}} (a+b+1-2 
p))}},\,\,\,\,\,\,\,\,\,a,b\,<\,l-1\,\, ,\cr
&S_{l a}(\theta)\,=\,\prod_{p=0}^{a-1}\,{{ 
F (\theta -{{i \pi}\over{h}} (l-a+2 p))}\over {
F (\theta +{{i \pi}\over{h}} (l-a+2 p))}}\,\, ,\cr
&S_{l l}(\theta)\,=\,\prod_{p=0}^{[{{l-2}\over{2}}]}\,{{ 
F (\theta -{{i \pi}\over{h}} (4 p+1))}\over {
F (\theta +{{i \pi}\over{h}} (4 p+1))}}\,\, ,\cr
&S_{l l-1}(\theta)\,=\,\prod_{p=1}^{[{{l-1}\over{2}}]}\,{{ 
F (\theta -{{i \pi}\over{h}} (4 p-1))}\over {
F (\theta +{{i \pi}\over{h}} (4 p-1))}}\,\, ,\cr
&S_{l-1 l-1}(\theta)\,=\,S_{l l}(\theta),\,\,\,\,\,\,\,\,\,
S_{l-1 a}(\theta)\,=\,S_{la}(\theta) \,\, .\cr}}
Here the symbol $[a]$ stands for integer part of $a$ and the rapidity variable 
$\theta$ is defined as 
\eqn\rapid{(p_{n}\,+\,p_{k})^{2}\,=\,m_{n}^{2}\,+\,m_{k}^{2}\,+
\,2 m_{n} m_{k} \cosh(\theta)\,\, .} 
The functions ${\cal F}(\theta)$ and $F(\theta)$ are given by
\eqn\deff{\eqalign{&{\cal 
F}(\theta)\,=\,{{\tanh({{\theta}\over{2}}+{{i \pi}\over{2 
h}}-{{i \pi}\over{h Q b}}) \,\tanh({{\theta}\over{2}}+{{i \pi}\over{2 
h}}-{{i \pi b}\over{h Q}})}\over{\tanh({{\theta}\over{2}}+{{i \pi}\over{2 
h}}) \,\tanh({{\theta}\over{2}}-{{i \pi}\over{2 
h}})}}\,\, ,\cr
&F (\theta)\,=\,{{\sinh({{\theta}\over{2}}+{{i \pi}\over{2 
h}}-{{i \pi}\over{h Q b}}) \,\sinh({{\theta}\over{2}}+{{i \pi}\over{2 
h}}-{{i \pi b}\over{h Q}})}\over{\sinh({{\theta}\over{2}}+{{i \pi}\over{2 
h}}) \,\sinh({{\theta}\over{2}}-{{i \pi}\over{2 
h}})}}\,\, .\cr}}
where $Q = b + b^{-1}$. 

In the physical strip $0<\,\Im m\,\theta<\,\pi$ the amplitudes\ \ampl \  
possess simple poles at $\theta = i \theta_{a b}^{c}$ corresponding 
to the bound states of particles $B_{a}$ and $B_{b}$. Specifically,
\item{(i)}  the amplitude $S_{l l}(\theta)$ has simple poles at
\eqn\polone{\theta\,=\,{{2 i \pi (2 k+1)}\over{h}},\,\,\,\,\,\,\,\,\,
k\,=0,\ldots,\,[{{l-1}\over{2}}]-1}
which represent particles $B_{l-2-2 k}$;
\item{(ii)} the amplitude $S_{l l-1}(\theta)$ has simple poles at
\eqn\poltwo{\theta\,=\,{{4 i \pi k}\over{h}},\,\,\,\,\,\,\,\,\,
k\,=1,\ldots,\,[{{l}\over{2}}]-1}
which represent particles $B_{l-3-2 k}$;
\item{(iii)} the amplitudes $S_{l a}(\theta)$ and $S_{l-1 a}(\theta)$  have a 
simple pole at 
\eqn\polthr{\theta\,=\,{{i \pi}\over{2}}\,+\,{{i \pi a}\over{h}}}
which represents particle $B_{l-1}$ or $B_{l}$ correspondingly; 
\item{(iv)} the amplitudes $S_{a c}(\theta)$, $a \le c$ have simple poles at
\eqn\polfou{\theta\,=\,{{i \pi (c + a)}\over{h}}, \hskip25pt 
\theta\,=\,i \pi\,-\,{{i \pi (c - a)}\over{h}} }
representing particles $B_{c + a}$ and $B_{c - a}$.

\noindent There are also  simple poles at $\theta\,=\,i 
\pi\,-\,i\,\theta_{a b}^{c}$. 
These poles correspond to the particles in the cross channel. The analytical structure of
the $S$-matrix respects the discrete $G$-symmetry \ \groupactionev ,\ \groupactionod \  of 
the model.

\newsec{Heuristic framework for  form factors}

Form factors are on-shell amplitudes of a local field ${\cal O}$
\eqn\form{F_{{\cal O}}(\theta_1,...\theta_n)=\langle\, vac\,|
\, \pi_A ({\cal O})\,  | 
\, B(\theta_1)...B(\theta_n)\, \rangle\ \, , }
where  the  matrix of the field \ ${\cal O}$\ in the basis of asymptotic states
is denoted by $\pi_A ({\cal O})$. The form factors satisfy a set of
requirements \fed \ which constitute a complicated 
Riemann-Hilbert problem.
In our calculation of form factors we will follow 
ideas of \sert \br \reshh . The main tool we will exploit is a special
representation $\pi_{Z}$ of the  formal Zamolodchikov-Faddeev  algebra associated 
with the $S$-matrix \ \ampl \ . Defining properties  of  $\pi_{Z}$ 
were discussed in \ser \br \ and were motivated by form factor axioms 
\fed . In particular, if we denote by ${\bf B}_{k}$ \ ZF operators acting in 
$\pi_{Z}$, they satisfy exchange relations
\eqn\exchrel{{\bf B}_{k}(\theta_1) {\bf B}_{p}(\theta_2)=
S_{k p}(\theta_1-\theta_2)\  {\bf B}_{p}(\theta_2) {\bf B}_{k}(\theta_1)\
,\hskip30pt \Im m\, (\theta_1-\theta_2)\,=\,0 \,\, .}
It is also assumed that there exists an operator \ ${\bf K}$\ acting  in the
space $\pi_Z$ in the following manner
\eqn\shift{{\bf B}_{k}(\theta+\alpha)=
e^{-\alpha {\bf K}}\, {\bf B}_{k}(\theta)\, e^{\alpha {\bf K}}\ .}
Unitarity and crossing symmetry of the $S$-matrix allow us to equip 
$\pi_{Z}$ with a conjugation operation 
\eqn\ksisyuty{{\bf B}^{+}_{k}(\theta)=
{\bf B}_{k}(\theta+i\pi)\ ,\hskip30pt {\bf K}^+=
-{\bf K} \ .}
There exists an embedding of the linear space of asymptotic
states $\pi_{A}$  in the tensor product  of $\pi_Z$ and its dual $\bar \pi_Z$,
\eqn\hsyt{\pi_A \,\hookrightarrow \, {\bar \pi_Z}\otimes \pi_Z\ .}
In other words, we can identify  an arbitrary  vector
$|\, X \,  \rangle\in \pi_A$\ with
some endomorphism  (linear operator) ${\bf  X}$ of
the space\ $\pi_Z$.
To describe the embedding,  we identify an arbitrary vector $|\,
B_{k_{1}}(\theta_1)...B_{k_{n}}(\theta_n)\, \rangle\in\pi_A$ with an element of
${\rm End}\big[\pi_Z\big]$ as 
\eqn\lssoij{|\, B_{k_{1}}(\theta_1)...B_{k_{n}}(\theta_n)\, \rangle \equiv
{\bf B}_{k_{1}}(\theta_1)...{\bf B}_{k_{n}}(\theta_n)\  e^{i \pi  {\bf K}}\ .}
The  asymptotic states generate a basis
in\ $\pi_A$, therefore\ \lssoij\ unambiguously specifies the
embedding of the linear space. As well as\ $\pi_A$,\ the
space\ ${\bar \pi_Z}\otimes \pi_Z$\ possesses  a  canonical 
Hilbert space structure with the scalar product given by
$${\rm Tr}_{\pi_Z}\Big[\,
{\bf Y}^+{\bf X}\, \Big]
/{\rm Tr}_{\pi_Z}\Big[ e^{2 \pi i  {\bf K}}\, \Big]\ .$$
The conjecture that allows effective calculation of form factors is 
that  the embedding \ \lssoij \ of the linear space $\pi_A \,\hookrightarrow \, {\bar
\pi_Z}\otimes \pi_Z$, preserves  the structure of the  Hilbert spaces,
\eqn\jsusyt{\langle\, Y\, |\, X \,  \rangle={\rm Tr}_{\pi_Z}\Big[\,
{\bf Y}^+{\bf X}\, \Big]
/{\rm Tr}_{\pi_Z}\Big[ e^{2 \pi i  {\bf K}}\, \Big]\, ,\ \ 
\ \ {\rm if}\ \ 
 |\, X \,  \rangle\equiv {\bf X}\, , \ \ 
|\, Y \,  \rangle\equiv {\bf Y}\ .}
Let us define $\pi_{Z}({\cal O})\in {\rm End}\big[\pi_Z\big]$ 
 associated with the state 
$\pi_{A}({\cal O})|\, vac \, \rangle$ in the following way
\eqn\mvnbb{ \pi_{A}({\cal O})|\, vac \, \rangle
\equiv  \pi_{Z}({\cal O}) 
\ e^{\pi i {\bf K}}\ .}
We also require that $\pi_{Z}({\cal O})\in {\rm End}
\big[\pi_{Z}\big]$ associated  with a local Hermitian field
must commute with ${\bf B}(\theta)$,
\eqn\lor{\big[\, \pi_{Z}({\cal O})\,  ,\, 
 {\bf B}_{k}(\theta)\, \big]=0\ .}
Using \ \jsusyt\ the form-factors can be written  as traces
over the space $\pi_Z$,
\eqn\traaaa{F_{{\cal O}}(\theta_1,...\theta_n)=
{\rm Tr}_{\pi_Z}\Big[\, e^{2\pi i {\bf K}}
\ \pi_{Z}({\cal O})\ {\bf B}_{k_{1}}(\theta_1)...{\bf B}_{k_{n}}(\theta_n)\, \Big]
/{\rm Tr}_{\pi_Z}\Big[ e^{2 \pi i {\bf  K}}\, \Big]\ .}

\newsec{Free field representation for form factors}

To construct the representation $\pi_{Z}$ for $D_{l}^{(1)}$ ATQFT consider a set 
of oscillators 
\eqn\set{ \lambda_{\nu}^{(a)} \,\, , \hskip20pt a\, \in \,J\,=\, \{1, \ldots ,l,{\bar l},
 \ldots , {\bar 1} \}\,\, . }
We specify an order in the set J as
\eqn\order{{\bar 1}\, \succ \cdots \succ \,{\overline 
{l-1}}\,\succ  \,{{\bar l}\atop 
{l}} \, \succ \,l-1\, \succ \cdots \succ\,1\,\, .}
These  oscillators satisfy commutation relations
\eqn\comoscone{\eqalign{&[ \lambda_{\nu}^{(a)},\lambda_{\nu'}^{(a)} ]\,=\,
{ {4 \cosh \bigl( \pi \nu ({1 \over 2}-{1 \over h}) \bigr) } \over
{\nu \cosh({{\pi \nu}\over2})}}\sinh ( {{\pi b \nu} 
\over {h Q}}) \sinh( {{\pi \nu} \over {h Q b}})\,\, 
\delta_{\nu+\nu',0}\, ,\cr
&[\lambda_{\nu}^{(\bar a)},\lambda_{\nu'}^{(a)} ]\,=\,
- 4 \nu^{-1} \, \bigg\{ {{e^{{{\pi \nu}\over {2}}} 
\sinh({{\pi \nu}\over h}) } \over {\cosh({{\pi 
\nu}\over2})}}\,+\,e^{{{\pi \nu}\over {h}} (2 l -2 a 
-1)} \bigg\}\sinh ( {{\pi b \nu} 
\over {h Q}}) \sinh( {{\pi \nu} \over {h Q b}}) \,\, 
\delta_{\nu+\nu',0}\, ,\cr
&[\lambda_{\nu}^{(a)},\lambda_{\nu'}^{(b)} ]\,=\,
- 4 \, \epsilon (a,b) \, {{e^{{{\epsilon(a,b) \pi \nu}\over {2}}} 
\sinh({{\pi \nu}\over h}) } \over {\nu \cosh({{\pi 
\nu}\over2})}}\,\sinh ( {{\pi b \nu} 
\over {h Q}}) \sinh( {{\pi \nu} \over {h Q b}}) \,\, 
\delta_{\nu+\nu',0}\, ,\hskip15pt b \,\ne \,{\bar a},\, a\cr}}
The function $\epsilon(a,b)$ is a step function:  $\epsilon(a,b)\,=\,-1$ 
if $a \prec b$ or $(a,b) = (l,{\bar l})$ , 
and  $\epsilon(a,b)\,=\,1$ if $a \succ b$ or $(a,b) = ({\bar l}, l)$ . 
The ``reflected''  oscillators $\lambda_{\nu}^{(\bar a)}$ are not 
independent and can be expressed in terms  of $\lambda_{\nu}^{(a)}$ ,
\eqn\refosc{\lambda_{\nu}^{(\bar a)}\,=\,-\lambda_{\nu}^{(a)} e^{{2 
\pi (l-a) \nu}\over h}\,-\,2 \sinh({{\pi \nu}\over h}) \, 
\sum_{k=1}^{a-1} \, \lambda_{\nu}^{(k)} e^{{\pi (2 l-2 k-1) \nu}\over 
h}\,.}

It is convenient to extend the Heisenberg algebra \  
\comoscone \ by canonical conjugate pairs of operators ${\bf P}^{(a)}$, ${\bf Q}^{(a)}$ 
commuting with the oscillators $\lambda_{\nu}^{(a)}$,
\eqn\canpair{\eqalign{&[{\bf P}^{(a)}, {\bf Q}^{(b)}]\,=\,-i 
\,(\delta_{a,b}\,-\,\delta_{a-1,b})\, , \hskip30pt b < n\, ,\cr
&[{\bf P}^{(a)}, {\bf Q}^{(n)}]\,=\,-i 
\,(\delta_{a,n}\,+\,\delta_{a-1,n})\, .}}
The extended algebra admits a representation in the direct sum of Fock 
spaces,
\eqn\fock{\pi_{Z}\,=\,\oplus_{{\vec p}} {\cal F}_{{\vec p}}\ ,\ \  {\rm where }
\ \ {\bf {\vec P}} {\cal F}_{\vec p}= {\vec p}\, {\cal F}_{\vec p}\ .}
and each of the spaces ${\cal F}_{\vec p}$ is a span 
\eqn\span{{\cal F}_{\vec p}:\   \oplus\,  
\lambda^{(a_{1})}_{-\nu_1}...\lambda^{(a_{n})}_{-\nu_n}|\, {\vec p}\, \rangle \,  ,\ \  \
\nu_k>0\ .}
The highest vector\ $|\, {\vec p}\, \rangle$
(not to be confused with the physical vacuum\ $ |\, vac\, \rangle$ of 
the model) obeys the equations \
$\lambda^{(a)}_{\nu} |\, {\vec p}\, \rangle\,=\,0\, ,\,\,\,  \nu>0.$ 
Let $\vec \rho $ be a half sum of the positive roots of the Lie 
algebra $D_{l}$,
\eqn\ro{{\vec \rho}\,{\vec \alpha}_{k}\,=\,1\,\, , \hskip30pt k\,=\,1,\ldots ,l\, ,}
and ${\vec h}_{a}$, $a \in J$ are the weights of of the first fundamental 
representation $\pi_{1}$
\eqn\weightsone{\eqalign{&\vec h_{a}\,=\,{\vec \varepsilon_{a}}\, ,\hskip30pt a = 
1,\ldots, l\, ,\cr 
&{\vec h}_{\bar a}\,=\,-{\vec \varepsilon}_{a}\, .\cr}}
Now we can define vertex operators \sert \fei  ,
\eqn\verla{\Lambda_{{a}} (\theta)\,=\,\exp\bigl({{2 i 
\pi}\over{h}}\,(\vec \rho\,-\,{{\bf {\vec P}}\over{Q}}) \cdot{\vec 
h}_{a}\,\bigr)\,:\exp\bigl(-\,i 
\int\limits_{-\infty}^{+\infty} d\nu \,\lambda_{{\nu}}^{(a)} e^{i \nu 
(\theta -{{i \pi}\over 2})}\,\bigr) :}
which are in one-to-one correspondence with the weights of 
the representation $\pi_{1}$ and satisfy exchange relations 
\eqn\lacom{\Lambda_{{a}} (\theta_{1}) \, \Lambda_{{b}} (\theta_{2})\,=\,
S_{{11}}(\theta_{1}-\theta_{2}) \,\Lambda_{{b}} (\theta_{2}) \, 
\Lambda_{{a}} (\theta_{1})\, , \hskip30pt \Im m\, 
(\theta_1-\theta_2)\,=\,0 \, .}
The operators $\Lambda_{{a}} (\theta)$ will be used to build ZF 
operators $\{{\bf B}_{k}(\theta)\}_{k=1}^{l-2}$ associated  with the vectorial representations 
$\{\pi_{k}\}_{k=1}^{l-2}$. The operators ${\bf 
B}_{l-1}(\theta)$, ${\bf B}_{l}(\theta)$ associated with spinorial representations 
$\pi_{l-1}$, $\pi_{l}$ cannot be expressed solely in terms of 
$\Lambda_{{a}} (\theta)$`s and we need to introduce additional vertex 
operators $A_{k}(\theta)$ and $Y_{k}(\theta)$ \reshh \ corresponding to the simple positive
roots ${\vec \alpha}_{k}$ and the highest weight vectors $\vec \omega_{k}$ of the fundamental
representations $\pi_{k}$ of the Lie algebra $D_{l}$. Consider new 
oscillators
\eqn\rootosc{\eqalign{&a_{\nu}^{(p)}\,=\,(\lambda_{\nu}^{(p)} - 
\lambda_{\nu}^{(p)}) e^{-{{\pi p \nu}\over{h}}}\, , \hskip30pt 
p\,=\,1,\ldots,l-1\,\, ,\cr
&a_{\nu}^{(l)}\,=\,(\lambda_{\nu}^{(l-1)}-  
\lambda_{\nu}^{(\bar l)}) e^{-{{\pi (l-1) \nu}\over{h}}}\, . \cr}}
and 
\eqn\weiosc{\eqalign{&y_{\nu}^{(p)}\,=\,\sum_{k=1}^{p}\,\lambda_{\nu}^{(k)} 
e^{{{\pi (p - 2 k+1) \nu}\over{h}}}\, , \hskip30pt p\,=\,1,\ldots,l-2\,\, ,\cr
&y_{\nu}^{(l-1)}\,=\, \bigl( 2 \cosh({{\pi \nu}\over{h}}) \bigr)^{-1}\, 
\bigl( \,\lambda_{\nu}^{(1)} 
e^{{{\pi (l-3) \nu}\over{h}}}\,+ \cdots +\,\lambda_{\nu}^{(l-1)} 
e^{-{{\pi (l-1) \nu}\over{h}}} \,-\,\lambda_{\nu}^{(l)} 
e^{-{{\pi (l-1) \nu}\over{h}}}\,\bigr)\, , \cr
&y_{\nu}^{(l)}\,=\, \bigl( 2 \cosh({{\pi \nu}\over{h}}) \bigr)^{-1}\, 
\bigl( \lambda_{\nu}^{(1)} 
e^{{{\pi (l-1) \nu}\over{h}}}\,+ \cdots +\,\lambda_{\nu}^{(l-1)} 
e^{-{{\pi (l-3) \nu}\over{h}}} \,+\,\lambda_{\nu}^{(l)} 
e^{-{{\pi (l-1) \nu}\over{h}}}\,\bigr)\, . \cr}}
The vertex operators $A_{k}(\theta)$ and $Y_{k}(\theta)$ are defined 
analogously to \ \verla 
\eqn\veray{\eqalign{&A_{{k}} (\theta)\,=\,\exp\bigl({{2 i 
\pi}\over{h}}\,(\vec \rho\,-\,{{\bf {\vec P}}\over{Q}}) \cdot{\vec 
\alpha}_{k}\,\bigr)\,
:\exp\bigl(-\,i \int\limits_{-\infty}^{+\infty} d\nu \,a_{{\nu}}^{(k)} e^{i \nu 
(\theta -{{i \pi}\over 2})}\,\bigr) :\,\, ,\cr
&Y_{{k}} (\theta)\,=\,\exp\bigl({{2 i 
\pi}\over{h}}\,(\vec \rho\,-\,{{\bf {\vec P}}\over{Q}}) \cdot{\vec 
\omega}_{k}\,\bigr)\,
:\exp\bigl(-\,i \int\limits_{-\infty}^{+\infty} d\nu \,y_{{\nu}}^{(k)} e^{i \nu 
(\theta -{{i \pi}\over 2})}\,\bigr) :\,\, .\cr}}
where the highest weight vectors  ${\vec \omega}_{p}$ are 
\eqn\ome{\eqalign{&{\vec \omega}_{p}\,=\,\sum_{k=1}^{p}\,{\vec 
\varepsilon}_{k}\,\, ,\hskip30pt p\,=\,1,\ldots,l-2\,\, ,\cr
&{\vec \omega}_{l-1}\,=\,{1 \over 2} ({\vec \varepsilon}_{1}\,+ 
\cdots +\,{\vec \varepsilon}_{l-1}\,-\,{\vec \varepsilon}_{l})\,\, ,\cr
&{\vec \omega}_{l}\,=\,{1 \over 2} ({\vec \varepsilon}_{1}\,+ 
\cdots +\,{\vec \varepsilon}_{l-1}\,+\,{\vec \varepsilon}_{l})\,\, 
.\cr}}
Using results from Appendix A one can show that for $\Im m\, (\theta_1-\theta_2)\,=\,0$
\eqn\aycom{\eqalign{&A_{{b}} (\theta_{1}) \, A_{{c}} (\theta_{2})\,=\,
A_{{c}} (\theta_{2}) \, A_{{b}} (\theta_{1})\,\, ,\cr 
&A_{{b}} (\theta_{1}) \, Y_{{c}} (\theta_{2})\,=\,
Y_{{c}} (\theta_{2}) \, A_{{b}} (\theta_{1})\,\, ,\cr
&Y_{{b}} (\theta_{1}) \, Y_{{c}} (\theta_{2})\,=\,S_{b 
c}(\theta_{1}-\theta_{2})\,
Y_{{c}} (\theta_{2}) \, Y_{{b}} (\theta_{1})\,\, .\cr}}

Now we proceed to the construction of ZF operators ${\bf B}_{k}(\theta)$. It was
mentioned in the Introduction that the free field representations of ZF 
operators are similar to  the  Baxter  $T\,-\,Q$ equations \bax . This observation is very 
useful in finding explicit forms of ${\bf B}_{k}(\theta)$'s.
Unfortunately, it seems that in the case of $D_{l}^{(1)}$ ATQFT 
the reconstruction of ZF operators from the Baxter equations is not 
straightforward. The general form of ${\bf B}_{k}(\theta)$ remains obscure 
and the best we can do is to use an operator product expansion to 
generate ZF operators step by step\foot{Another possible rout is to 
use a recurrent procedure of \pilch \ to generate currents of deformed 
$W(D_{l}^{(1)})$ algebra. Then, as it was proposed in \br , one might 
hope to recover ZF operators by taking a ``scaling'' limit of the currents. }.

First, following \reshh ,  define ${\bf B}_{1}(\theta)$,
\eqn\bone{{\bf B}_{1}(\theta)\,=\,Q \sqrt{{{h \kappa_{1}}\over{2 
\pi}}}\,\sum_{a \in J}\,\Lambda_{a}(\theta)\,\, .}
where  $\kappa_{1}$ is some constant. Obviously, due to  \ 
\lacom \ ,
\eqn\bcom{ {\bf B}_{1}(\theta_{1}) \,{\bf B}_{1}(\theta_{2})\,=\,
S_{{11}}(\theta_{1}-\theta_{2})\,
{\bf B}_{1}(\theta_{2}) \,{\bf B}_{1}(\theta_{1})\,\, .}
We expect that operator products ${\bf B}_{p}(\theta_{1}) {\bf 
B}_{1}(\theta_{2})$ develop simple  poles in accordance with \ \polfou \
\eqn\opekone{{\bf B}_{p}(\theta_{1}) \,{\bf B}_{1}(\theta_{2}) 
\,\rightarrow\, {{{\bf B}_{p+1}(\theta_{2} + {{i \pi}\over{h}} )}\over
{\theta_{1}-\theta_{2}-{{i \pi (p+1)}\over{h}}}}\,\, , 
\hskip30pt \theta_{2}\,\rightarrow\ \,\theta_{1}-{{i \pi (p+1)}\over{h}}\,\, .}
After some calculations one 
can find the next few operators. For example,
\eqn\bs{\eqalign{{\bf B}_{2}&(\theta)\,=\,Q \sqrt{{{h \kappa_{2}}\over{2 
\pi}}}\,\bigg\{\sum_{\{a_{1}, a_{2}\} \in I,}\,\gamma_{a_{1} a_{2}} \,
\Lambda_{a_{1}} \Lambda_{a_{2}}\,\,+\,
 \gamma\,\Lambda_{\overline{l-2}}\, \Lambda_{l-2}^{\prime}\,-\,\gamma\,
\Lambda_{\overline{l-1}}\, \Lambda_{l-1}^{\prime} \,\bigg\}\,\, .\cr}}
The $\kappa_{p}$'s are constants which at this point are irrelevant.
The summation in \ \bs \ extends over the standard set $I$. We 
will call a set $\{a_{1},\ldots,a_{n}\}$, $a_{k} \in J $ and a monom 
$\Lambda_{a_{1}} \cdots  \Lambda_{a_{k}}$ standard 
if for any $k$ either $a_{k}\succ a_{k+1}$ or 
$(a_{k},a_{k+1})\,=\,({\bar l},l)\,\, or \,\,(l,{\bar l})$ . All monoms in \ \bs \   
depend on $\theta$  in same way, for example 
\eqn\thedep{\eqalign{&\Lambda_{a_{1}}
\Lambda_{a_{2}}^{\prime}\,\equiv\, \Lambda_{a_{1}}(\theta + {{i \pi} \over{h}}) \,
\Lambda_{a_{2}}^{\prime}(\theta - {{i \pi}\over{h}})\,\, 
.\cr}} 
The numerical coefficients $\gamma_{a_{1} a_{2}}$ and $c$ are found to 
be 
\eqn\pairconst{\eqalign{&\gamma_{{\overline{a}}{a}}\,=\,{{\bigl(l - 1 - a 
-{b \over Q}\bigr) \,\bigl(l - 1 - a -{1 \over {Q b}}\bigr)}\over{ 
\bigl(l - 1 - a \bigr) \,\bigl(l - 2 - a \bigr)}}\,\, ,
\hskip15pt a\, \ne \, l-1, l-2\,\, ,\cr
&\gamma_{{\overline{l-2}} \,{l-2}}\,=\,\gamma_{{\overline{l-1}}\,{l-1}}\,=\,1 - 
{{1}\over{Q^{2}}}\,\, ,\hskip20pt
\gamma_{l{\overline{l}}}\,=\,\gamma_{{\overline{l}}l}\,\, ,\cr
&\gamma\,=\,{{2 i \pi}\over{h Q^{2}}}\,\, .\cr}}
and $\gamma_{a b}\,=\,1$ for the rest of the cases. 
The definition of $\gamma_{{\overline{l-2}} {l-2}}$ and 
$\gamma_{{\overline{l-1}} {l-1}}$ has an ambiguity. As a matter of 
fact some of the monoms can coincide due to \ \refosc . For example,
\eqn\coila{ \Lambda_{\overline{l-p-1}}(\theta + {{i \pi p}\over{h}}) 
\Lambda_{l-p-1}(\theta - {{i \pi p}\over{h}})\,=\,
\Lambda_{\overline{l-p}}(\theta + {{i \pi p}\over{h}}) 
\Lambda_{l-p}(\theta - {{i \pi p}\over{h}})\,\, .}
Therefore, only the sum  $\gamma_{{\overline{l-2}} {l-2}} +  
\gamma_{{\overline{l-1}} {l-1}}$ enters the definition of ${\bf 
B}_{2}(\theta)$ . We choose 
$\gamma_{{\overline{l-2}} {l-2}} =  \gamma_{{\overline{l-1}} {l-1}}$ 
for convenience.  The identities \ \coila \ and similar identities 
allow also to cancel second and  higher order poles in the operator product 
expansions of  ${\bf B}_{k}$'s. The explicit form of the operator ${\bf 
B}_{3}(\theta)$ and the discussion of a general form of ZF operators are 
given in  Appendix B. 

The operators ${\bf B}_{l-1}(\theta)$ and ${\bf B}_{l}(\theta)$ 
cannot be obtained by the bootstrap procedure described above because 
the corresponding particles never appear as bound states of the particles 
$B_{k}$, $k \,=\, 1,\ldots ,l-2$. The weights of the spinorial representations $\pi_{l-1}$, 
$\pi_{l}$ of the Lie algebra $D_{l}$ are non-degenerate . Therefore, we expect, as 
in the case of ${\bf B}_{1}(\theta)$, that the 
form of ${\bf B}_{l-1}(\theta)$ and ${\bf B}_{l}(\theta)$ is 
essentially the same as of the corresponding $T\,-\,Q$ equation \ku  
\resh . 
For  $D_{4}^{(1)}$ ATQFT $(h = 6)$ we obtain
\eqn\spinb{\eqalign{&{\bf B}_{3}(\theta)\,=\,Q \sqrt{{{h \kappa_{3}}\over{2 
\pi}}}\,\cdot\cr
&\bigg\{Y_{3}(\theta)\,+: Y_{3}(\theta) A_{3}^{-1} (\theta - {{i 
\pi}\over{h}}) :\,+\cr
&: Y_{3}(\theta) A_{3}^{-1} (\theta - {{i \pi}\over{h}}) 
A_{2}^{-1} (\theta - {{2 i \pi}\over{h}}) :\,+\cr
&: Y_{3}(\theta) A_{3}^{-1} (\theta - {{i \pi}\over{h}})
 A_{2}^{-1} (\theta - {{2 i \pi}\over{h}}) A_{4}^{-1} (\theta - {{3 i 
 \pi}\over{h}}) :\,+\cr
&: Y_{3}(\theta) A_{3}^{-1} (\theta - {{i \pi}\over{h}}) 
A_{2}^{-1} (\theta - {{2 i \pi}\over{h}}) A_{1}^{-1} (\theta - {{3 i 
\pi}\over{h}}) :\,+\cr
&: Y_{3}(\theta) A_{3}^{-1} (\theta - {{i \pi}\over{h}}) 
A_{2}^{-1} (\theta - {{2 i \pi}\over{h}}) A_{1}^{-1} (\theta - {{3 i \pi}\over{h}})
 A_{4}^{-1} (\theta - {{3 i \pi}\over{h}}) :\,+\cr}}
\eqn\spinbcon{\eqalign{ 
 &: Y_{3}(\theta) A_{3}^{-1} (\theta - {{i \pi}\over{h}}) 
A_{2}^{-1} (\theta - {{2 i \pi}\over{h}}) A_{1}^{-1} (\theta - {{3 i \pi}\over{h}})
 A_{4}^{-1} (\theta - {{3 i \pi}\over{h}}) A_{2}^{-1} (\theta - {{4 i 
 \pi}\over{h}}) :\,+\cr
&: Y_{3}(\theta) A_{3}^{-1} (\theta - {{i \pi}\over{h}}) 
A_{2}^{-1} (\theta - {{2 i \pi}\over{h}}) A_{1}^{-1} (\theta - {{3 i \pi}\over{h}})
 A_{4}^{-1} (\theta - {{3 i \pi}\over{h}}) A_{2}^{-1} (\theta - {{4 i 
 \pi}\over{h}})\,\cdot \cr
&\hskip265pt \cdot  A_{3}^{-1} (\theta - {{5 i \pi}\over{h}}) : \bigg\}\cr}}
The operator ${\bf B}_{4}(\theta)$ can be obtained 
from \ \spinb \ by interchange of indices $3 \leftrightarrow 4$. To 
build ${\bf B}_{l-1}(\theta)$ and ${\bf B}_{l}(\theta)$ for general 
$D_{l}^{(1)}$ ATQFT one should use either the  recurrent procedure of \ku \ or 
a general formula from \resh . 

In the following sections we are going to deal with one- and two-particle 
form factors of an exponential operator
\eqn\expop{{\cal O}\,=\,e^{{\vec a} {\vec \varphi }}\,\, .}
To this end we need to specify the endomorphism $\pi_{Z}(e^{{\vec a} {\vec \varphi }})\in 
{\rm End}\big[\pi_Z\big]$   \ \mvnbb . It was observed in \sert \br \ that 
that for exponential operators the proper endomorphism is a projector 
on the Fock space ${\cal F}_{\vec a}$ with a given value of ``zero modes``
\eqn\projsp{{\vec {\bf P}} {\cal F}_{\vec a}\,=\,{\vec a}\, {\cal F}_{\vec 
a}\,\, .}
We will see that this assertion is also true for $D_{l}^{(1)}$ ATQFT. 
With this choice of $\pi_{Z}(e^{{\vec a} {\vec \varphi }})$  we have 
\eqn\trace{\langle\, vac\, |\, e^{{\vec a} {\vec \varphi }} \, |\,
B(\theta_1)\ldots B(\theta_n)\, \rangle \,=\,\langle \, e^{{\vec a} {\vec \varphi }} \,
 \rangle \,{\rm Tr}_{{\cal F}_{a}}\Big[\, e^{2\pi i {\bf K}}\, {\bf 
B}(\theta_1)\ldots {\bf B}(\theta_n)\, \Big]/{\rm Tr}_{{\cal F}_{a}}
\Big[ e^{2 \pi i {\bf  K}}\, \Big]\ \,\, ,}
where 
\eqn\onepoint{\langle \, {e^{{\vec a} {\vec \varphi }}} \, \rangle\,=\,
{\rm Tr}_{{\cal F}_{a}}\Big[ e^{2 \pi i  {\bf  K}}\, \Big]/{\rm Tr}_{\pi_Z}\Big[ e^{2 \pi i  {\bf  K}}
\, \Big]\ \,\, }
is the one-point function of the exponential operator which was 
calculated in \fat . 

Before we go to the general case we will present results for the first 
nontrivial case - $D_{4}^{(1)}$ ATQFT. This will also provide a test 
for the free field representation constructed in this chapter.

\newsec{$D_{4}^{(1)}$ Toda model}

For any simple Lie algebra one can define characters 
$\chi_{\omega_{p}}({\vec\lambda})$ of the $p$-fundamental representation 
$\pi_{p}$ by the formula 
\eqn\charp{\chi_{\omega_{p}}({\vec \lambda})\,=\,{\rm Tr}_{\pi_{p}}\, \Big[ 
e^{{{2 i \pi}\over{h}} ({\vec \rho}-{\vec \lambda}) {\vec H}} 
\Big]\,\, .}
Here ${\vec H}\,=\, (H_{1},\ldots ,H_{l})$ is a basis in the Cartan 
subalgebra of $D_{l}$, normalized with respect to the Killing form 
$\langle ,\rangle $ , $\langle H_{a},H_{b} \rangle\,=\,\delta_{a b}$.

Using the free field representation given in the last section we 
immediately derive the one-particle form factors,
\eqn\onedfou{\langle\, vac\, |\, e^{{\vec a} {\vec \varphi }} \, |\,
B_{p}(\theta)\, \rangle \,=\,\langle \, e^{{\vec a} {\vec \varphi }} \, 
\rangle \,\, Q \sqrt{{{3 Z_{p}}\over{\pi}}}\,{\cal X}_{p} 
\bigl({{\vec a}\over{Q}}\bigr) \,\, .}
The functions ${\cal X}_{p}$ can be expressed in terms of the the 
characters of $D_{4}$
\eqn\calx{\eqalign{&{\cal 
X}_{p}({\vec\lambda})\,=\,\chi_{\omega_{p}}({\vec\lambda})\,\, ,\hskip40pt p = 
1, 3, 4\,\,\cr
&{\cal X}_{2}({\vec\lambda})\,=\,\chi_{\omega_{2}}({\vec\lambda})\,+\,1\,\, 
.\cr}}
In the next section we will show that the constants $Z_{p}$ are the wave functions
renormalization constants and calculate them for general $D_{l}^{(1)}$ ATQFT. 

The calculation of two-particle form factors is also straightforward. In particular, 
we find form factors involving the particle $B_{1}$
\eqn\twodfou{\eqalign{
\langle\, vac\, |\, e^{{\vec a} {\vec \varphi }} \, |\,
B_{1}(\theta_{1}) B_{3}(\theta_{2})\, \rangle \,=&\, \langle \, e^{{\vec a} {\vec \varphi }} 
\, \rangle \,\,{{3 Q^{2}}\over{\pi}}\,{\sqrt {Z_{1} Z_{3}}}\, {\cal 
R}_{13}(\theta)\,\bigl(\,{\cal X}_{4}\, {\cal K}_{4} (\theta) + {\cal X}_{1} {\cal X}_{3}
\,\bigr)\,\, ,\cr
\langle\, vac\, |\, e^{{\vec a} {\vec \varphi }} \, |\,
B_{1}(\theta_{1}) B_{1}(\theta_{2})\, \rangle \,=&\,\langle \, e^{{\vec a} {\vec \varphi }} \, 
\rangle \,\,{{3 Q^{2}}\over{\pi}}\,Z_{1}\, {\cal R}_{11} 
(\theta) \,\bigl(\,{\cal X}_{2}\, {\cal K}_{2} 
(\theta) + {\cal X}_{1}^{2}\,\bigr)\,\, ,\cr
\langle\, vac\, |\, e^{{\vec a} {\vec \varphi }} \, |\,
B_{1}(\theta_{1}) B_{2}(\theta_{2})\, \rangle \,=&\, \langle \, e^{{\vec a} {\vec \varphi }} 
\, \rangle \,\,{{3 Q^{2}}\over{\pi}}\,{\sqrt {Z_{1} Z_{2}}}\, {\cal 
R}_{12}(\theta)\,\cdot \cr
&\bigl(\,{\tilde {\cal X}_{3}}\, {\cal K}_{3} 
(\theta) + {\cal X}_{1} {\cal X}_{2} - (1 - \eta){\cal 
X}_{1}\, {\cal K}_{1} (i \pi - \theta) \,\bigr)\,\, .\cr}}
The functions ${\cal R}_{1 p}(\theta) , \, p =1,2,3$ are  the ``minimal`` 
form  factors \karow . They admit an integral representation
\eqn\minform{\eqalign{&{\cal R}_{1 p}(\theta)\,=\cr
&\exp\bigg\{-2 \int\limits_{-\infty}^{\infty}{{d\nu}\over{\nu}}\,e^{i \nu 
(\theta - i \pi)}\, {{\sinh({{\pi b \nu}\over{h Q}}) \,\sinh({{\pi 
\nu}\over{h Q b}}) \,\cosh \bigl(\pi 
\nu ({{1}\over{2}} - {{p}\over{h}} ) \bigr)}\over
{\sinh(\pi \nu)\, \cosh({{\pi \nu}\over{2}})}}\,\bigg\}\,\, .\cr}}
The representation \ \minform \ is valid in the strip
\eqn\analstrip{- {{\pi (p -1)}\over{h}}\,<\,\Im 
m\,\theta\,<\, 2 \pi + {{\pi (p -1)}\over{h}}\,\, ,}
while outside the 
strip it must be understood in a sense of analytical continuation.  
In \ \minform \ , \analstrip \ $h=6$. 
The constant $\eta$, the ``character`` ${\tilde {\cal X}_{3}}$ 
and the functions ${\cal K}_{p}(\theta)$ are given by
\eqn\newco{\eqalign{
&{\tilde {\cal X}_{3}}\,=\,\chi_{\omega_{3} + \omega_{4}} + ( 1 + 
\eta)\,\chi_{\omega_{1}}\,\, ,\cr
&\eta\,=\,{4 \over{\sqrt 3}}\,\sinh({{i 
\pi}\over{6 Q b}}) \, \sinh({{i \pi 
b}\over{6 Q}})\,\, ,\cr
&{\cal K}_{p}(\theta)\,=\,
-\,{{2 i\,\sinh({{i \pi}\over{6 Q b}}) \,\sinh({{i \pi b}\over{6 Q}}) 
\,\sinh({{i \pi p}\over{6}})}
\over{\sinh({{\theta}\over{2}}+{{i \pi p}\over{12}}) 
\,\sinh({{\theta}\over{2}}-{{i \pi p}\over{12}})}}\,\, .\cr}}
The $D_{4}^{(1)}$ Toda model possesses a symmetry under the action of 
the permutation group $G = S_{3}$. The group $G$ permutes particles 
$B_{1}, B_{3}, B_{4}$ or , equivalently, it acts as a permutation on 
the set of weights $\{\omega_{1}, \omega_{3}, \omega_{4}\}$. We can use 
this symmetry to  obtain the rest of the form factors except for $\langle\, 
vac\, |\,  e^{{\vec a} {\vec \varphi }} \, |\,
B_{2}(\theta_{1}) B_{2}(\theta_{2})\, \rangle$. The expression for it 
is somewhat complicated and we do not present it here. 
 
One can show that for $j = 0, \ldots ,4$
\eqn\charray{\eqalign{\chi_{\omega_{1}}(\lambda {\vec \alpha}_{j})\,=&\,8\,\sinh({{i 
\pi \lambda}\over{6}})\,\sinh({{i \pi (1-\lambda)}\over{6}})\,\cosh({{i 
\pi (2 j-1)}\over{6}})\,\, , \cr
\chi_{\omega_{2}}(\lambda {\vec \alpha}_{j})\,=&\,8 {\sqrt 3}\,\sinh({{i 
\pi \lambda}\over{6}})\,\sinh({{i \pi (1-\lambda)}\over{6}}) 
\,\cosh({{i \pi (2 j-1)}\over{3}})\, +\cr
&16\,\sinh^{2}({{i \pi \lambda}\over{6}})\,\sinh^{2}({{i \pi 
(1-\lambda)}\over{6}})\,-1\,\, ,\cr}}
and 
\eqn\charrays{\eqalign{\chi_{\omega_{3}}(\lambda {\vec 
\alpha}_{0})\,=&\,-\,\chi_{\omega_{3}}(\lambda {\vec \alpha}_{1})\,=\,
-\,\chi_{\omega_{3}}(\lambda {\vec \alpha}_{3})\,=\,\chi_{\omega_{3}}(\lambda 
{\vec \alpha}_{4})\,=\cr
\chi_{\omega_{4}}(\lambda {\vec 
\alpha}_{0})\,=&\,-\,\chi_{\omega_{4}}(\lambda {\vec \alpha}_{1})\,=\,
\,\chi_{\omega_{4}}(\lambda {\vec \alpha}_{3})\,=-\,\chi_{\omega_{4}}(\lambda 
{\vec \alpha}_{4})\,=\cr
&4 {\sqrt {3}}\,\sinh({{i \pi \lambda}\over{6}})\,\sinh({{i 
\pi (1-\lambda)}\over{6}})\,\, , \cr
\chi_{\omega_{3}}(\lambda {\vec 
\alpha}_{2})\,=&\,\,\chi_{\omega_{4}}(\lambda {\vec 
\alpha}_{2})\,=\,0\,\, , \cr}}
This  allows to check that all one- and two-particle form factors  satisfy 
the quantum equations  of motion
\eqn\eqmoone{\partial_{\mu} \partial^{\mu} ({\vec \alpha}_{j} {\vec 
\varphi})\,=\,{\cal M}^{2}\, \sum_{k=1}^{l}\,{\bf C}_{j k} n_{k} \big( e^{b{\vec 
\alpha}_{k} {\vec \varphi}}\,-\,e^{b {\vec \alpha}_{0} {\vec 
\varphi}}\big)\,\, .}
with
\eqn\massdfour{{\cal M}^{2}\,=\,{{\pi\, m^{2}\, \sin({{\pi}\over{h}}) 
}\over{2 h O Q \sin({{\pi 
b}\over{h Q}})\, \sin({{\pi}\over{h Q b}})}}\,\, ,}
and $O\,=\,\langle e^{b {\vec \alpha_{j}} {\vec \varphi} } \rangle$ , 
$j\,=\,0,\ldots,l$, \ \ $h = 6$ . We intentionally keep the 
general notation for the dual Coxeter number $h$ in \ \minform  , \massdfour \ 
because, as we will see in the next section, the formulae are valid 
for general $h$.

Note also that the two-particle form factors exhibit the cluster 
property \delsica 
\eqn\claster{\langle\, vac\, |\, e^{{\vec a} {\vec \varphi }} \, |\,
B_{k}(\theta_{1}) B_{p}(\theta_{2})\, \rangle \,\rightarrow \,{{\langle\, vac\, |\, e^{{\vec a} {\vec \varphi }} \, |\,
B_{k}\, \rangle \,\langle\, vac\, |\, e^{{\vec a} {\vec \varphi }} \, |\,
B_{p}\, \rangle \,}\over{\langle\, e^{{\vec a} {\vec \varphi }} \rangle\,}}\,\, ,}
as $|\theta_{1} - \theta_{2}| \rightarrow \infty $.

\newsec{The wave function renormalization constants}

To find the wave function renormalization constants we need an 
explicit form of one- and two-particle form factors.
 Although  a general 
form of ZF operators is not known, it is still possible to 
carry out calculations building ${\bf B}_{k}(\theta)$ one by one 
and analyzing  the general case using formulae from Appendices B and 
C.  The results can  be  summarized as follows. 

\subsec{The one-particle form factors}
 
The one-particle form factors for $D_{l}^{(1)}$ ATQFT are given by
\eqn\onegen{\langle\, vac\, |\, e^{{\vec a} {\vec \varphi }} \, |\,
B_{p}(\theta)\, \rangle \,=\,\langle \, e^{{\vec a} {\vec \varphi }} \, 
\rangle \,\, Q \sqrt{{{h Z_{p}}\over{2 \pi}}}\,{\cal X}_{p} 
\bigl({{\vec a}\over{Q}}\bigr) \,\, .}
The functions ${\cal X}_{p}$ can be expressed in terms of the the 
characters of $D_{l}$,
\eqn\calxgen{\eqalign{&{\cal 
X}_{p}({\vec\lambda})\,=\,\sum_{s=0}^{[{{p}\over{2}}]} \zeta_{p\, p-2s} 
\,\chi_{\omega_{p - 2 s}}({\vec\lambda})\,\, ,\hskip40pt p = 
1, \ldots  l-2\,\, ,\cr
&{\cal X}_{p}({\vec\lambda})\,=\, \chi_{\omega_{p}}({\vec\lambda})\,\, ,\hskip40pt p = 
l-1, l\,\, .\cr}}
Unfortunately, the constants $\zeta_{p k}$ cannot be calculated in a 
closed form. Instead, we found the following representation for them,
\eqn\zeco{\zeta_{p k}\,=\,1 + \xi_{p\, p-2} + \xi_{p\, p-4} + \cdots + 
\xi_{p k}\,\, ,}
where $\xi_{p k}$ is obtained by recursion,
\eqn\recurs{\xi_{p k}\,=\,{{\big[{{b}\over{Q}}\big]_{x}\, 
\big[{{1}\over{Q b}}\big]_{x}}\over{[1]_{x}}}
\,\sum_{s=1}^{{{p-k}\over{2}}}\,\xi_{p\,  k+2s} \, 
{{[k]_{x}\, [k+s]_{x}\, [s]_{x}}\over{[p-k]_{x}\,[p+k]_{x}}} \,\, .}
Here a notation $[\,\cdot\,]_{x}$ was introduced, 
\eqn\conv{[a]_{x}\,=\,x^{a}-x^{-a}\,\, , \hskip40pt 
x = {{i \pi}\over{h}}\,\, .}
In the above formulae we set $\xi_{p p} \equiv 1$ and one can also 
find that $\xi_{p 0} \equiv 0$ . In particular it means that  
$\zeta_{p 2} = \zeta_{p 0}$ .

Taking a limit of small ${\vec a}$ in \ \onegen \ we find the one-particle 
form factors of the field ${\vec \varphi}$ itself,
\eqn\figen{\langle\, vac\, |\, {\vec \alpha}_{j} {\vec \varphi } \, |\,
B_{p}(\theta)\, \rangle \,= \,- \, \sqrt{{{2 \pi 
Z_{p}}\over{h}}}\,{\aleph}_{p} (j) \,\, .}
The functions ${\aleph}_{p}(j)$ are found to be
\eqn\alef{\eqalign{&{\aleph}_{p}(j)\,=\,-i\,\sum_{k=1}^{p} \xi_{p k} \,
{{[k]_{x}\,[2 k\, (2 j - 1)]_{x}}\over{[k \,(2 j - 1)]_{x}}}\,\, 
,\hskip40pt p = 1,\ldots,l-2\,\, ,\cr}}
and ${\aleph}_{l-1}(j)$, ${\aleph}_{l}(j)$ take a form
\eqn\alesp{\eqalign{&{\aleph}_{l-1}(0)\,=\,-\,{\aleph}_{l-1}(1)\,=\,- 
i^{-l}\,{\aleph}_{l-1}(l-1)\,=\,i^{-l}\,{\aleph}_{l-1}(l)\,=\,{\sqrt { 
{{h}\over {2}}}}\,\, ,  \cr
&{\aleph}_{l}(0)\,=\,-\,{\aleph}_{l}(1)\,=\, 
i^{-l}\,{\aleph}_{l}(l-1)\,=\,-i^{-l}\,{\aleph}_{l}(l)\,=\,
{\sqrt { {{h}\over {2}}}}\,\, ,\cr
&{\aleph}_{l-1}(j)\,=\,{\aleph}_{l}(j)\,\equiv\,0\,\, ,\hskip70pt 
j=2,\ldots,l-2\,\, .\cr}}
In the course of derivation of \ \alef  ,\alesp \ we used the 
generating function,
\eqn\gener{G(t,j,\lambda)\,=\,(1-e^{2 
t})\,{{[t+j-\lambda]_{x}\,[t-j+\lambda]_{x}\,[t+j-1+\lambda]_{x}\,[t-j+1-\lambda]_{x}}
\over{[t+j]_{x}\,[t-j]_{x}\,[t+j-1]_{x}\,[t-j+1]_{x}}}\,\, .}
The first $l-2$ terms of the expansion of $G(t,j,\lambda)$,
\eqn\defgen{G(t,j,\lambda)\,=\,\sum_{s=0}^{\infty}\,e^{s t} \chi_{\omega_{s}}(\lambda {\vec 
\alpha}_{j})\,\, .}
give the values of  $\chi_{\omega_{s}}(\lambda {\vec \alpha}_{j})$ for 
$s = 0,\ldots ,l-2$. The values of  $\chi_{\omega_{l-1}}(\lambda {\vec 
\alpha}_{j})$ and $\chi_{\omega_{l}}(\lambda {\vec \alpha}_{j})$ can 
be found explicitly,
\eqn\spin{\eqalign{&\chi_{\omega_{l-1}}(\lambda {\vec 
\alpha}_{0})\,=\,-\chi_{\omega_{l-1}}(\lambda {\vec \alpha}_{1})\,=\,- 
i^{-l}\,\chi_{\omega_{l-1}}(\lambda {\vec 
\alpha}_{l-1})\,=\,i^{-l}\,\chi_{\omega_{l-1}}(\lambda {\vec 
\alpha}_{l})\,=\,\,  \cr
&\chi_{\omega_{l}}(\lambda {\vec 
\alpha}_{0})\,=\,-\,\chi_{\omega_{l}}(\lambda {\vec \alpha}_{1})\,=\, 
i^{-l}\,\chi_{\omega_{l}}(\lambda {\vec 
\alpha}_{l-1})\,=\,-i^{-l}\,\chi_{\omega_{l}}(\lambda {\vec 
\alpha}_{l})\,=\cr
&\hskip55pt i {\sqrt {2 h}}\,\,{{\sinh({{i \pi \lambda}\over{h}})\,\sinh({{i 
\pi (1-\lambda)}\over{h}})}\over{\sinh({{i \pi}\over{h}})}}\,\, ,\cr
&\chi_{\omega_{l-1}}(\lambda {\vec 
\alpha}_{j})\,=\,\chi_{\omega_{l}}(\lambda {\vec 
\alpha}_{j})\,\equiv\,0\,\, ,\hskip40pt j=2,\ldots,l-2\,\, .\cr}}
One can show, using \ \gener \  and \ \spin , that the one-particle form 
factors  satisfy the quantum  equations of motion \ \eqmoone \ for general 
$h$. 

The fields $ \varphi^{(a)}$ are normalized in accordance with the short 
distance behavior, 
\eqn\limcor{\langle\, vac\, |\, 
\varphi^{(a)}(x)\,\varphi^{(b)}(y)\,|\, vac\, \rangle \,=\, -2 
\delta^{a b}\,\log ( M |x-y| ) + O(1)\,\, , \hskip20pt |x-y|\, 
\rightarrow\, 0\,\, .}
This normalization and \ \newfieldev \ - \newfieldod \ , \figen \ 
allow to identify $Z_{p}$ in \ \onegen \ as the wave function 
renormalization constants and to find  one-particle form factors 
of the field $ \vec \phi $
\eqn\matfi{\eqalign{&\langle\,vac\,|\,\phi^{(k)}\,|\,B_{p}\,\rangle \,=\, 
{\sqrt {2 \,\pi  Z_{p}}}\,\xi_{ p k}\,\, ,\hskip30pt p = 1,\ldots 
,l-2\,\, ,\cr
&\langle\,vac\,|\,\phi^{(k)}\,|\,B_{p}\,\rangle \,=\, 
{\sqrt {2 \,\pi  Z_{p}}}\,\delta_{ p k}\,\, ,\hskip30pt p = l-1, l\,\, .}}

\subsec{Two-particle form factors}

For the purpose of calculating $Z_{p}$ it is enough to calculate 
the simplest two-particle form factors 
\eqn\twogen{\eqalign{\langle\, vac\, |\, e^{{\vec a} {\vec \varphi }} \, |&\,
B_{1}(\theta_{1}) B_{1}(\theta_{2})\, \rangle \,=\, \langle \, e^{{\vec a} {\vec \varphi }} 
\, \rangle \,\,{{h Q^{2}}\over{2 \pi}}\,Z_{1}^{2}\, {\cal 
R}_{1 1}(\theta)\, \bigl(\,{\cal X}_{2}\, {\cal K}_{2} 
(\theta) + {\cal X}_{1}^{2}\,\bigr)\,\, ,\cr
\langle\, vac\, |\, e^{{\vec a} {\vec \varphi }} \, |&\,
B_{1}(\theta_{1}) B_{p}(\theta_{2})\, \rangle \,=\, \langle \, e^{{\vec a} {\vec \varphi }} 
\, \rangle \,\,{{h Q^{2}}\over{2 \pi}}\,{\sqrt {Z_{1} Z_{p}}}\, {\cal 
R}_{1 p}(\theta)\,\cdot \cr
&\bigl(\,{\cal X}_{p+1}\, {\cal K}_{p+1} 
(\theta) + {\cal X}_{1} {\cal X}_{p} - \eta_{p} {\cal 
X}_{p-1}\, {\cal K}_{p-1} (i \pi - \theta) \,\bigr)\,\, , \hskip20pt 
p = 2, \ldots ,l-3\,\, ,\cr
\langle\, vac\, |\, e^{{\vec a} {\vec \varphi }} \, |&\,
B_{1}(\theta_{1}) B_{l-2}(\theta_{2})\, \rangle \,=\, \langle \, e^{{\vec a} {\vec \varphi }} 
\, \rangle \,\,{{h Q^{2}}\over{2 \pi}}\,{\sqrt {Z_{1} Z_{l-2}}}\, {\cal 
R}_{1 l-2}(\theta)\,\cdot \cr
&\bigl(\,{\tilde {\cal X}}_{l-1}\, {\cal K}_{l-1} 
(\theta) + {\cal X}_{1} {\cal X}_{l-2} - \eta_{l-2} {\cal 
X}_{l-3}\, {\cal K}_{l-3} (i \pi - \theta) \,\bigr)\,\, ,\cr
\langle\, vac\, |\, e^{{\vec a} {\vec \varphi }} \, |&\,
B_{1}(\theta_{1}) B_{l-1}(\theta_{2})\, \rangle \,=\, \langle \, e^{{\vec a} {\vec \varphi }} 
\, \rangle \,\,{{h Q^{2}}\over{2 \pi}}\,{\sqrt {Z_{1} Z_{l-1}}}\, {\cal 
R}_{1 l-1}(\theta)\, \bigl(\,{\cal X}_{l}\, {\cal K}_{l} 
(\theta) + {\cal X}_{1} {\cal X}_{l-1}\,\bigr)\,\, ,\cr
\langle\, vac\, |\, e^{{\vec a} {\vec \varphi }} \, |&\,
B_{1}(\theta_{1}) B_{l}(\theta_{2})\, \rangle \,=\, \langle \, e^{{\vec a} {\vec \varphi }} 
\, \rangle \,\,{{h Q^{2}}\over{2 \pi}}\,{\sqrt {Z_{1} Z_{l}}}\, {\cal 
R}_{1 l}(\theta)\, \bigl(\,{\cal X}_{l-1}\, {\cal K}_{l} 
(\theta) + {\cal X}_{1}{\cal X}_{l}\,\bigr)\,\, .\cr}}
and form factors involving both of the particles $B_{l-1}$ and $B_{l}$
\eqn\spintwof{\eqalign{\langle\, &vac\, |\, e^{{\vec a} {\vec \varphi }} \, |\,
B_{l-1}(\theta_{1}) B_{l-1}(\theta_{2})\, \rangle \,=\,\cr
&\langle \, e^{{\vec a} {\vec \varphi }} 
\, \rangle \,\,{{h Q^{2}}\over{2 \pi}}\,Z_{l-1}\, {\cal 
R}_{l-1 l-1}(\theta)\, \Bigl(\,\sum_{p=1}^{[{{l-1}\over 
2}]}\,\bigl(\,\prod_{k=1}^{p-1} \eta_{l-1-2 k}\bigr)\,{\cal X}_{l-2 
p}\,{\cal K}_{4 p - 2}(\theta)\,+\,{\cal X}_{l-1}^{2} \,\Bigr)\,\, 
,\cr
\langle\, &vac\, |\, e^{{\vec a} {\vec \varphi }} \, |\,
B_{l}(\theta_{1}) B_{l}(\theta_{2})\, \rangle \,=\,\cr
&\langle \, e^{{\vec a} {\vec \varphi }} 
\, \rangle \,\,{{h Q^{2}}\over{2 \pi}}\,Z_{l}\, {\cal 
R}_{l l}(\theta)\, \Bigl(\,\sum_{p=1}^{[{{l-1}\over 
2}]}\,\bigl(\,\prod_{k=1}^{p-1} \eta_{l-1-2 k}\bigr)\,{\cal X}_{l - 2 
p}\,{\cal K}_{4 p-2}(\theta)\,+\,{\cal X}_{l}^{2} \,\Bigr)\,\, ,\cr
\langle\,& vac\, |\, e^{{\vec a} {\vec \varphi }} \, |\,
B_{l-1}(\theta_{1}) B_{l}(\theta_{2})\, \rangle \,=\,\cr
&\langle \, e^{{\vec a} {\vec \varphi }} 
\, \rangle \,\,{{h Q^{2}}\over{2 \pi}}\,{\sqrt{Z_{l-1} Z_{l}}}\, {\cal 
R}_{l-1 l}(\theta)\, \Bigl(\,\sum_{p=1}^{[{{l-2}\over 
2}]}\,\bigl(\,\prod_{k=1}^{p-1} \eta_{l-1-2 k}\bigr)\,{\cal X}_{l-1-2 
p}\,{\cal K}_{4 p}(\theta)\,+\,{\cal X}_{l-1}{\cal X}_{l} \,\Bigr)\,\, 
.\cr}}
where the functions ${\cal K}_{p}(\theta)$, the constants 
${\eta}_{p}$ and the ``character`` ${\tilde {\cal 
X}}_{l-1}({\vec\lambda})$ are similar to the ones introduced 
in the case of the $D_{4}^{(1)}$ Toda model \ \newco \
\eqn\mumuo{\eqalign{&
{\tilde {\cal 
X}}_{l-1}({\vec\lambda})\,=\,\chi_{\omega_{l}+\omega_{l-1}}({\vec\lambda}) + 
\,\sum_{s=1}^{[{{l-1}\over{2}}]} \zeta_{p\, p-2s} 
\,\chi_{\omega_{p - 2 s}}({\vec\lambda})\,\, ,\cr
&{\eta}_{p}\,\,=\,{{\,\big[l - 1 - p -{{1}\over{Q b}}\big]_{x}
\,\big[l - 1 - p -{{b}\over{Q}}\big]_{x}}\over{[l-1-p]_{x} 
\,[l-p]_{x}}}\,\,  ,\cr
&{\cal K}_{p}(\theta)\,\,=\,{{[ p ]_{x}}\over{4 [1]_{x}}}\,{{\,\big[{{1}\over{Q b}}\big]_{x} 
\,\big[{{b}\over{Q}}\big]_{x}}
\over{\sinh({{\theta}\over{2}}+{{i \pi p}\over{2 h}}) 
\,\sinh({{\theta}\over{2}}-{{i \pi p}\over{2 h}})}}\,\, .\cr}}
The ``minimal`` form factors  ${\cal R}_{1 p}(\theta) ,\,\,p = 
1,\ldots ,l-1$, ${\cal R}_{1 l-1}(\theta)\,=\,{\cal 
R}_{1 l}(\theta)$ are given by \ \minform \ with general $h$. 
For ${\cal R}_{l-1 l}(\theta)$ and ${\cal R}_{l-1 l-1}(\theta)\,=\,{\cal 
R}_{l l}(\theta)$ we have
\eqn\spinfor{\eqalign{&{\cal R}_{l-1 l}(\theta)\,=\,
\exp\bigg\{- \int\limits_{-\infty}^{\infty}{{d\nu}\over{\nu}}\,e^{i \nu 
(\theta - i \pi)}\, {{\sinh({{\pi b \nu}\over{h Q}}) \,\sinh({{\pi 
\nu}\over{h Q b}}) \,\sinh({{\pi \nu (l-2)}\over{h}})}\over
{\sinh(\pi \nu)\, \cosh({{\pi \nu}\over{2}})\,\sinh({{2 \pi 
\nu}\over{h}})}}\,\bigg\}\,\, ,\cr
&{\cal R}_{l l}(\theta)\,=\,
\exp\bigg\{- \int\limits_{-\infty}^{\infty}{{d\nu}\over{\nu}}\,e^{i \nu 
(\theta - i \pi)}\, {{\sinh({{\pi b \nu}\over{h Q}}) \,\sinh({{\pi 
\nu}\over{h Q b}}) \,\sinh({{\pi \nu l}\over{h}})}\over
{\sinh(\pi \nu)\, \cosh({{\pi \nu}\over{2}})\,\sinh({{2 \pi 
\nu}\over{h}})}}\,\bigg\}\,\, .\cr}}

The dependence on the vector 
$\vec a$ enters the form factors through the ``characters`` ${\cal 
X}_{p}\big({{{\vec a}}\over Q}\big)$. 

\subsec{The calculation of the renormalization constants $Z_{p}$}

The form factors \ \twogen , \spintwof \ satisfy crossing symmetry and 
the Watson 
equations \fed \ and exhibit the required analytical structure. In particular, 
they have simple poles at the points $\theta\,=\,i\,\theta_{a b}^{c}$  \ \polthr \ , \polfou \  which 
correspond to bound states of particles. The singularity 
of the form factors
\eqn\formfacsing{\langle\, vac\, |\, e^{{\vec a} {\vec \varphi }} \, |\,
B_{a}(\theta_{1}) B_{b}(\theta_{2})\, \rangle \,\rightarrow \, {{i 
\Gamma_{a b}^{c}}\over{\theta_{1} - \theta_{2} - i\,\theta_{1 
a}^{b}}}\,\langle \, vac\, | \, e^{{\vec a} {\vec \varphi }} \, | \, 
B_{c}\, \rangle \,\, ,\hskip20pt  \theta\,\rightarrow \, 
\,i \, \theta_{a b}^{c}\,\, ,}
defines \fed \ the residue of the two-particle $S$-matrix
\eqn\ressmat{S_{a b}(\theta)\,\to\,{ {i\,(\Gamma^{c}_{a b})^2}
\over{\theta\,-\,i\,\theta_{1 a}^{b}}}\,\,  , \hskip30pt  \theta\,\rightarrow \, 
\,i \, \theta_{a b}^{c}\,\, .} 
Comparison of $ \Gamma_{a b}^{c} $ obtained from the $S$-matrix and 
the form factor singularities leads to the equations
\eqn\zeq{\eqalign{&{{Z_{1} Z_{p}}\over{Z_{p+1}}}\,=\, {{4 i \pi}\over{h 
Q^{2}}} \, {{[2]_{x}}\over{\big[{{2 b}\over{ 
Q}}\big]_{x}\,\big[{{2}\over{Q b}}\big]_{x}}}\,{{{\cal F} (i \pi-{{i \pi (2 p + 
1)}\over{h}})}\over {{\cal R}^{2}_{1 p} ({{i \pi (p + 
1)}\over{h}})}}\,\, ,\cr
&{{Z_{1} Z_{p}}\over{Z_{p-1}}}\,=\, {{4 i \pi}\over{h 
Q^{2}}} \, {{[2]_{x}}\over{\big[{{2 b}\over{ 
Q}}\big]_{x}\,\big[{{2}\over{Q b}}\big]_{x}}}\,{{{\cal F} (i \pi-{{i \pi (2 p - 
1)}\over{h}})}\over {\eta_{p}^{2}\,{\cal R}^{2}_{1 p} ({{i \pi (p - 
1)}\over{h}})}}\,\, ,\cr
&Z_{l}^{2}\,=\,
{{2 i \pi}\over{h Q^{2}}} \, {{[1]_{x}}\over{\big[{{ b}\over{ 
Q}}\big]_{x}\,\big[{{1}\over{Q b}}\big]_{x}}}\,{{Z_{l-2 p}}\over {{\cal 
R}^{2}_{ll}  ({{i \pi (4 p - 
2)}\over{h}})}}\,
{{\prod\limits_{k=0}^{p-2} F({{i \pi (4 p - 4 k 
- 3)}\over{h}})\,\prod\limits_{k=p}^{[{{l-2}\over2}]} F({{i \pi (4 k - 4 p 
+ 3)}\over{h}})}\over{ \prod\limits_{k=1}^{p-1}\eta_{l-1-2 k}^{2} 
\,\prod\limits_{k=0}^{[{{l-2}\over2}]} F({{i \pi (4 p + 4 k - 1)}\over{h}}) 
}}\,\, .\cr}}
which unambiguously determine  the constants $Z_{p}$,
\eqn\ofuckwegotitfinally{\eqalign{&\hskip5pt Z_{p}\,=\cr
&\exp \bigg\{ - 4 \int\limits_{0}^{\infty} {{d \nu} \over {\nu}}    
\, \sinh({{\pi b \nu}\over{h Q}}) \sinh({{\pi 
\nu}\over{h Q b}}) \,\Bigl( {{\cosh( \pi 
\nu)\,\sinh({{\pi  p \nu}\over{h}})\,\cosh \bigl( \pi 
\nu ({{1}\over{2}} -
{{p}\over{h}}) \bigr) }\over
{\sinh(\pi \nu)\, \sinh({{\pi \nu}\over{h}})\,\cosh({{\pi 
\nu}\over{2}})}} - e^{-{{\pi \nu}\over{h}}}\Bigl) \bigg\}\,\, ,\cr}}
for $p = 1,\ldots, l-2$, and $Z_{l-1} =  Z_{l}$
\eqn\fuckwegotitfinally{\eqalign{&\hskip5pt Z_{l}\,=\cr
&\exp \bigg\{ - 2   \int\limits_{0}^{\infty} {{d \nu} \over {\nu}}    
\, \sinh({{\pi b \nu}\over{h Q}}) \sinh({{\pi 
\nu}\over{h Q b}}) \,\Bigl( {{\cosh( \pi 
\nu)\,\sinh({{\pi l \nu}\over{h}})}\over
{\sinh(\pi \nu)\, \sinh({{2 \pi \nu}\over{h}})\,\cosh({{\pi 
\nu}\over{2}})}} - e^{-{{\pi \nu}\over{h}}}\Bigl) \bigg\}\,\, .\cr}}
The above  formulae coincide with \ \mainf \ if we recall the explicit form of 
$C^{-1}$ \reshh . The exact result  for $Z_{p}$ \ 
\ofuckwegotitfinally ,\fuckwegotitfinally \
also matches the one loop perturbative check.

\newsec{Conclusion}

In this paper we have calculated the wave function renormalization 
constants. The formula \ \mainf , together with the 
formulae for the one-particle form factors \ \onegen \ may be considered as 
the main results of the paper. Although the absence of a general expression 
for  ZF generators ${\rm B}_{k}(\theta)$ acting in $\pi_{Z}$ makes 
the analysis of multi-particle form factors difficult, it is still possible 
to obtain two-particle form factors involving the particle  
$B_{1}$ and the particles $B_{l-1}, B_{l}$ in an asymptotic state. 

It would be interesting (if possible)  to find the explicit form of ZF 
operators or a generating function for them. Another interesting problem 
is to find one-particle form factors for other Toda field theories and 
to understand their group-theoretical meaning. Note, that in the limit 
$b \rightarrow 0$ the one-particle form factors of $D_{l}^{(1)}$, as 
well as of $A_{l}^{(1)}$, ATQFT become the characters of 
finite dimensional representations of the Yangian $Y(D_{l})$ or 
$Y(A_{l})$ correspondingly \fate . And, finally, we consider as a challenging 
problem to generalize the  result \ \mainf \ to Toda theories associated 
with non-simply laced Lie algebras.

\bigskip

\centerline{\bf Acknowledgments}

I am grateful to S. Lukyanov for interesting discussions. I also 
would like to thank R. Nepomechie and L. Mezincescu for helpful 
discussions and the Department of Physics and Astronomy, Rutgers 
University for hospitality. This work was supported in part by the 
National  Science Foundation under Grants PHY-9507829 and PHY-9870101.

\newsec{Appendix A}

Here we collect commutation relations and pairings  of the operators 
$\Lambda_{{k}} (\theta)$, $A_{{k}} (\theta)$ and $Y_{{k}} (\theta)$. 
In all formulae below we denote $\theta\,=\,\theta_{1}-\theta_{2}$.

\eqn\pairingone{ \eqalign{ &\langle \,{\vec p}\,|\,\Lambda_{{a}} (\theta_{1}) \Lambda_{{a}} 
(\theta_{2})\,|\,{\vec p}\,\rangle\,=\,g(\theta)\,e^{{{4 i 
\pi}\over{h}}\,(\vec \rho\,-\,{{\vec p}\over{Q}}) {\vec 
h}_{a}}\,\, ,\cr
&\langle\,{\vec p}\,|\,\Lambda_{{a}} (\theta_{1}) { \Lambda_{{\bar a}} 
(\theta_{2})}\,|\,{\vec p}\,\rangle\,=\,g(\theta) \,f(\theta +{{i \pi}\over h}) 
f(\theta +{{i \pi (2 l-2 a-1)}\over h})\,\, ,\cr
&\langle\,{\vec p}\,|\,{\Lambda_{{\bar a}} (\theta_{1})} {\Lambda_{{a}} 
(\theta_{2})}\,|\,{\vec p}\,\rangle\,=\,g(\theta) \, f(\theta -{{i \pi}\over h}) 
f(\theta -{{i \pi (2 l-2 a-1)}\over h}) \,\, , \cr}}
and for $b\,\ne \,a,\,{\bar a}$ we obtain
\eqn\pairingtwo{\langle\,{\vec p}\,|\,\Lambda_{{a}} (\theta_{1}) \Lambda_{{b}} 
(\theta_{2})\,|\,{\vec p}\,\rangle\,=\,g(\theta) \, 
f(\theta - {{i \pi \epsilon(a,b)}\over h})\,e^{{{2 i 
\pi}\over{h}}\,(\vec \rho\,-\,{{\vec p}\over{Q}}) ({\vec 
h}_{a}+{\vec h}_{b}) }\,\, ,}
where the function $g(\theta)$ is defined for $\Im m\, \theta\,\geq \,0$ by
\eqn\defgf{\eqalign{&g(\theta)\,=\,\exp \bigg\{\,-4 
\int\limits_{0}^{\infty}{{d\nu}\over{\nu}}\,e^{i \nu \theta}\,
\sinh ( {{\pi b \nu} \over {h Q}}) \, \sinh( {{\pi \nu} \over {h Q 
b}})\,{{\cosh \bigl( \pi \nu ({1 \over 2}-{1 \over h}) \bigr) } \over {
\cosh({{\pi \nu}\over2})}}\,\bigg\}\, ,\cr
&f(\theta)\,=\,{{({{\theta}\over{2}}+{{i \pi}\over{2 
h}}-{{i \pi}\over{h Q b}}) \,({{\theta}\over{2}}+{{i \pi}\over{2 
h}}-{{i \pi b}\over{h Q}})}\over{({{\theta}\over{2}}+{{i \pi}\over{2 
h}}) \,({{\theta}\over{2}}-{{i \pi}\over{2 h}})}}\,\, .\cr}}
For $\Im m\, \theta\,<\,0$ relations \ \pairingone \ - \defgf \ 
must be understood in the  sense of analytical continuation.

The commutation relations 
\eqn\comay{\eqalign{&[ a_{\nu}^{(p)},a_{\nu'}^{(q)} ]\,=\,
4 \nu^{-1}\,C_{p q}\,\sinh ( {{\pi b \nu} 
\over {h Q}}) \sinh( {{\pi \nu} \over {h Q b}})\,\, 
\delta_{\nu+\nu',0}\,\, ,\cr
&[ y_{\nu}^{(p)},y_{\nu'}^{(q)} ]\,=\,
4 \nu^{-1}\,(C^{-1})_{p q}\,\sinh ( {{\pi b \nu} 
\over {h Q}}) \sinh( {{\pi \nu} \over {h Q b}})\,\, 
\delta_{\nu+\nu',0}\,\, ,\cr
&[ a_{\nu}^{(p)},y_{\nu'}^{(q)} ]\,=\,
4 \nu^{-1}\,\delta_{p q}\,\sinh ( {{\pi b \nu} 
\over {h Q}}) \sinh( {{\pi \nu} \over {h Q b}})\,\, 
\delta_{\nu+\nu',0}\,\, ,\cr}}
allows us to calculate pairings between $Y_{a}(\theta)$ and 
$A_{b}(\theta)$
\eqn\normay{\eqalign{&\langle \,{\vec p}\,|\, Y_{a}(\theta_{1}) 
Y_{b}(\theta_{2 })\,|\,{\vec p}\,\rangle\,=\,R_{a 
b}(\theta)\,e^{{{2 i 
\pi}\over{h}}\,(\vec \rho\,-\,{{\vec p}\over{Q}}) ({\vec 
\omega}_{a}+{\vec \omega}_{b})}\,\, ,\cr
&\langle \,{\vec p}\,|\,Y_{a}(\theta_{1}) A_{a}(\theta_{2 })\,|\,{\vec p}\,\rangle\,=\,f^{-1}(\theta) 
\,e^{{{2 i 
\pi}\over{h}}\,(\vec \rho\,-\,{{\vec p}\over{Q}}) ({\vec 
\omega}_{a}+{\vec \alpha}_{a}) }\,\, ,\cr
&\langle \,{\vec p}\,|\, A_{a}(\theta_{1}) A_{a}(\theta_{2 })\,|\,{\vec p}\,\rangle\,=\,f^{-1}(\theta + {{i 
\pi}\over{h}}) f^{-1}(\theta - {{i \pi}\over{h}})\,e^{{{4 i 
\pi}\over{h}}\,(\vec \rho\,-\,{{\vec p}\over{Q}}) {\vec 
\alpha}_{a} } \,\, ,\cr
&\langle \,{\vec p}\,|\, A_{a}(\theta_{1}) 
A_{b}(\theta_{2})\,|\,{\vec p}\,\rangle\,=\,f(\theta)\,e^{{{2 i 
\pi}\over{h}}\,(\vec \rho\,-\,{{\vec p}\over{Q}}) ({\vec 
\alpha}_{a}+{\vec \alpha}_{b}) }\,\, ,\hskip30pt  {\rm 
if}\,\,\,\,\,{\bf C}_{a 
b}\,=\,-1\,\, ,\cr}}
and the rest of the pairings are equal to $1$. The function $R_{a 
b}(\theta)$, $a \le b$ is given by
\eqn\rab{R_{a b}(\theta)\,=\,\exp \bigg\{\,-4 
\int\limits_{0}^{\infty}{{d\nu}\over{\nu}}\,e^{i \nu \theta}\, \sinh ( {{\pi b \nu} 
\over {h Q}})\,\sinh( {{\pi \nu} \over {h Q 
b}})\,\bigl(C^{-1}(\nu)\bigr)_{a 
b}\,\bigg\}\,\, .}
This integral representation is valid for $\Im m \theta \ge -{{\pi (b - 
a)}\over{h}}$ and must be understood in a sense of analytical continuation 
for  $\Im m \theta \le -{{\pi (b - a)}\over{h}}$

\newsec{Appendix B}

Here we give the explicit form of the operator $B_{p}(\theta)$  for 
$p = 3$ and discuss the case of general $p$.

\eqn\btre{\eqalign{&\hskip10pt{\bf B}_{3}(\theta)\,=\,Q \sqrt{{{h \kappa_{3}}\over{2 
\pi}}}\,\bigg\{\sum_{\{a_{1}, a_{2}, a_{3}\} \in 
I,}\,\gamma \bigl(\Lambda_{a_{1}}, \Lambda_{a_{2}}, \Lambda_{a_{3}}\bigr) \,
\Lambda_{a_{1}} \Lambda_{a_{2}} \Lambda_{a_{3}} \,\,+\cr
 &\gamma \sum_{a=1}^{l-3} \left( \Lambda_{\overline{l-2}}\, (\Lambda_{l-2} 
\Lambda_{a})^{\prime}\,+\,
\Lambda_{\overline a} \Lambda_{\overline{l-2}}
\Lambda_{l-2}^{\prime}\right)\,
 - \,\gamma \sum_{a=1}^{l-2} \left( \Lambda_{\overline{l-1}}\, (\Lambda_{l-1} 
\Lambda_{a})^{\prime}\,+ \, \Lambda_{\overline a}\, \Lambda_{\overline{l-1}} 
\Lambda_{l-1}^{\prime}\right)\,+\cr
&\gamma \sum_{a=l-2}^{l}\left( \Lambda_{\overline{l-3}}\, (\Lambda_{a} 
\Lambda_{l-3})^{\prime}\, + \,\Lambda_{\overline{l-3}} \Lambda_{\overline a}
\Lambda_{l-3}^{\prime}\right)\, -
\,\gamma \sum_{a=l-1}^{l}\left( \Lambda_{\overline{l-2}}\, (\Lambda_{a} 
\Lambda_{l-2})^{\prime}\,+\, \Lambda_{\overline{l-2}}\, 
\Lambda_{\overline a} \Lambda_{l-2}^{\prime}\right)\bigg\} \cr}}
The numerical coefficients $\gamma\,\bigl(\Lambda_{a_{1}}, \ldots , 
\Lambda_{a_{n}}\bigr)$ have a factorized form
\eqn\const{\eqalign{&\gamma \bigl(\Lambda_{a_{1}}, \ldots , 
\Lambda_{a_{p}}\bigr)\,=\,\prod_{m<k}\,\gamma  \bigl(\Lambda_{a_{m}},\,\, 
\Lambda_{a_{k}}\bigr)\,\, ,\cr}}
where 
\eqn\pairconst{\eqalign{&\gamma \bigl(\Lambda_{\overline{a}}(\theta + {{ 
2 i \pi p}\over{h}}),\,\, \Lambda_{a}(\theta) \bigr)\,=\,c_{a + p - 1}\,\, ,
\hskip15pt a+p\, \ne \, l, l-1\,\, ,\cr
&\gamma \bigl(\Lambda_{\overline{l-2}}(\theta + {{2 i \pi 
}\over{h}}),\,\, \Lambda_{l-2}(\theta) 
\bigr)\,=\,\gamma \bigl(\Lambda_{\overline{l-1}}(\theta +  
{{2 i \pi}\over{h}}),\,\, \Lambda_{l-1}(\theta) \bigr)\,=\,1 - 
{{1}\over{Q^{2}}}\,\, ,\cr
&\gamma \bigl(\Lambda_{l}\bigl(\theta + {{2 i \pi 
p}\over{h}}),\,\, \Lambda_{l}(\theta) \bigr)\,=\,\gamma \bigl( 
\Lambda_{\overline{l}}(\theta + {{2 i \pi 
p}\over{h}}),\,\, \Lambda_{\overline{l}}(\theta) 
\bigr)\,=\,c^{-1}_{l+ p - 2}\,\, ,\cr
&\gamma \bigl(\Lambda_{l}(\theta + {{2 i \pi 
p}\over{h}}),\,\, \Lambda_{\overline{l}}(\theta) \bigr)\,=\,\gamma 
\bigl(\Lambda_{\overline{l}}(\theta + {{2 i \pi 
p}\over{h}}),\,\, \Lambda_{l}(\theta) \bigr)\,=\,c_{l + p - 1}\,\, 
,\cr
&\gamma\,\bigl(\Lambda_{a},\,\, 
\Lambda_{b}\bigr)\,=\,1\,\, ,\hskip30pt {\rm 
for\,\,the\,\,rest\,\,of\,\,the\,\,cases}\,\, .\cr}}
The constants $c_{p}$ and $\gamma$ are  given by
\eqn\fuc{\eqalign{&c_{p}\,=\,1 + {{1}\over{(l - 1 - p) (l - 2 - p )\,
Q^{2}}}\,\, ,\cr
&\gamma\,=\,{{2 i \pi}\over{h Q^{2}}}\,\, .\cr}}
For general $p$ an operator ${\bf B}_{p}(\theta)$  can be written in 
a form
\eqn\bp{\eqalign{{\bf B}_{p}(\theta)\,=\,Q \sqrt{{{h \kappa_{p}}\over{2 
\pi}}}\,\bigg\{\sum_{\{a_{1}, \ldots , a_{p}\} \in 
I,}\,\gamma \bigl(\Lambda_{a_{1}}, \ldots , \Lambda_{a_{p}}\bigr) \,
\Lambda_{a_{1}} \cdots \Lambda_{a_{p}}\,+
{\rm ``unpleasant``\,\,terms}\,\bigg\}\,\, .\cr}}
The derivative  terms are not the only possible ``unpleasant'' terms 
that can appear in  ${\bf B}_{p}(\theta)$. Considering ${\bf 
B}_{4}(\theta)$ one can find, for example,  that 
besides derivative terms it contains also the nonstandard 
monom $\Lambda_{\overline{l-3}} 
\Lambda_{\overline{l-2}} \Lambda_{l-3} \Lambda_{l-2}$. All ``unpleasant''  
terms disappear  in the limit $b \rightarrow 0$ 
and $ {\bf B}_{p}(\theta)$ acquire a form of the Baxter $T\,-\,Q$ 
equation.

\newsec{Appendix C}

Here we give some details of the trace calculations. 

\noindent The problem is to calculate traces over a Fock space
\eqn\trace{\langle \langle\,{\cal O}(\lambda)\,\rangle \rangle 
\,=\,{{{\rm Tr}_{{\cal F}_{\vec p}}[\,e^{2 \pi i {\bf K}}\,{\cal 
O}(\lambda)\,]}\over{{\rm Tr}_{{\cal F}_{\vec p}}[\,e^{2 \pi i {\bf 
K}}\,]}}\,\, ,} 
where ${\cal O}$ is an arbitrary operator. This can be achieved by 
adopting a method of \clava \ . Let the oscillators  
$\lambda_{\nu}^{(a)}$ satisfy
\eqn\cool{[ \lambda_{\nu}^{(a)},\, \lambda_{\nu'}^{(b)} 
]\,=\,\nu^{-1}\,g^{a b}(\nu) \,\,\delta_{\nu + \nu',0}\,\, .}
We introduce a complementary set of 
oscillators $\gamma_{\nu}^{(a)}$ with commutation relations
\eqn\coog{\eqalign{&[ \gamma_{\nu}^{(a)},\, \gamma_{\nu'}^{(b)} 
]\,=\,\nu^{-1}\,g^{a b}(-\nu)\,\, \delta_{\nu + \nu',0}\,\, ,\cr
&[ \gamma_{\nu}^{(a)},\, \gamma_{\nu'}^{(b)} ]\,=\,0\,\, .\cr}}
Then the traces can be rewritten as vacuum averages 
\eqn\rewr{\eqalign{&\hskip10pt
{{\rm Tr}_{{\cal F}_{\vec p}}[\,e^{2 \pi i {\bf K}}\,{\cal 
O}(\lambda)\,]}\,= \cr
&\langle\, v \,|\, \exp 
\Bigl( \, \int\limits_{0}^{\infty}\,d\nu \nu \, g_{a b}(\nu)\, 
\gamma_{\nu}^{(a)} \lambda_{\nu}^{(b)} \,\Bigr)\,{\cal O}(\lambda) \, \exp 
\Bigl( \, \int\limits_{0}^{\infty}\,d\nu \nu \, g_{a b}(-\nu)\, 
\gamma_{-\nu}^{(a)} \lambda_{-\nu}^{(b)} \,\Bigr)\, |\, v \,\rangle \,\, 
.\cr}}
where 
\eqn\defgg{g_{a b}(\nu)\, g^{b c}(\nu)\,=\,\delta^{c}_{a}\,\, , 
\hskip40pt \gamma_{\nu}^{(a)} \, |\, v \,\rangle \,=\,\lambda_{\nu}^{(a)} 
\, |\, v \,\rangle \,=\, 0\,\, ,\,\,\,\,\,\nu >0\,\, .}
After some manipulations one can arrive at
\eqn\tran{\eqalign{&\hskip10pt {\rm Tr}_{{\cal F}_{\vec p}}[\,e^{2 \pi 
i {\bf K}} \,{\cal 
O}(\lambda_{-\nu}^{(a)},\, \lambda_{\nu}^{(b)})\,]\,=\cr
&\langle\, v \,|\,{\cal O} \bigl( \lambda_{-\nu}^{(a)} + {{e^{-2 \pi 
\nu}}\over{1-e^{-2 \pi \nu}}}\, \gamma_{\nu}^{(a)} ,\,\, 
\gamma_{-\nu}^{(a)} + {1 \over{1-e^{-2 \pi \nu}}}\, 
\lambda_{\nu}^{(a)}\,| \,v \,\rangle \,\, {\rm Tr}_{{\cal F}_{a}}[\,e^{2 
\pi i {\bf K}}\,]\,\, .\cr}}
It is implied in \ \tran \ that $\nu > 0$. Now one can find, for example
\eqn\trla{\eqalign{&\langle \langle 
\,\Lambda_{a}(\theta_{1}) \,\Lambda_{a}(\theta_{2}) \,\rangle \rangle \,=\,{\cal N}\,{\cal 
R}_{1 1}(\theta)\,e^{{{4 i 
\pi}\over{h}}\,(\vec \rho\,-\,{{\vec p}\over{Q}}) {\vec 
h}_{a}}\,\, ,\cr
&\langle \langle 
\,\Lambda_{a}(\theta_{1}) \,\Lambda_{\bar a}(\theta_{2}) \,\rangle 
\rangle \,=\,{\cal N}\,{\cal 
R}_{1 1}(\theta)\,F(\theta +{{i \pi}\over h})\, 
F(\theta +{{i \pi (2 l-2 a-1)}\over h})\,\, ,\cr
&\langle \langle 
\,\Lambda_{\bar a}(\theta_{1}) \,\Lambda_{a}(\theta_{2}) 
\,\rangle \rangle \,=\,{\cal N}\,{\cal R}_{1 1}(\theta) \, F(\theta -{{i \pi}\over h})\, 
F(\theta -{{i \pi (2 l-2 a-1)}\over h}) \,\, , \cr}}
and for $b\,\ne \,a,\,{\bar a}$ we obtain
\eqn\paingtwo{\langle \langle
\,\Lambda_{a}(\theta_{1}) \,\Lambda_{b}(\theta_{2}) \,\rangle \rangle \,=\,{\cal N}\,{\cal R}_{1 
1}(\theta) \, F(\theta-{{i \pi \epsilon(a,b)}\over h})\,e^{{{2 i 
\pi}\over{h}}\,(\vec \rho\,-\,{{\vec p}\over{Q}})  ({\vec 
h}_{a}+{\vec h}_{b}) }\,\, .}
The constant $\cal N$ can be absorbed in the definition of $Z_{p}$ and 
its  exact value is not important.

\vfill
\eject 

\listrefs
\end